\documentclass[aps,a4paper,superscriptaddress,showpacs,preprint,endfloats*,prl,longbibliography,floatfix]{revtex4-1} 
\usepackage{tabularx}
\usepackage{amsmath}
\usepackage{amssymb}
\usepackage{array}
\usepackage{multirow}
\usepackage{rotating}
\usepackage{subfigure}
\usepackage{graphicx}
\usepackage[english]{babel}

\makeatletter
\adddialect\l@ENGLISH\l@english
\makeatother

\begin{document}
\setlength{\tabcolsep}{1.5mm} 
\setcounter{totalnumber}{4}
\setcounter{topnumber}{4}
\setlength{\voffset}{-0cm}
\setlength{\hoffset}{-0.cm}
\addtolength{\textheight}{1.1cm}




\newcommand{\figwidth}{0.90\columnwidth}
\newcommand{\eq}[1]{Eq.(\ref{#1})}
\newcommand{\fig}[1]{Fig.~\ref{#1}}
\newcommand{\sect}[1]{Sec.~\ref{#1}}
\newcommand{\avg}[1]{{\langle #1 \rangle}}
\newcommand{\olcite}[1]{Ref.~\onlinecite{#1}}




\title{Ultrathin fibres from electrospinning experiments under driven fast-oscillating perturbations}
\author{Ivan Coluzza}
\email{i.coluzza@iac.cnr.it;ivan.coluzza@univie.ac.at}
\affiliation{CNR-Istituto per le Applicazioni del Calcolo ``Mauro Picone'', Via dei Taurini 19, I-00185 Rome, Italy}
\affiliation{Faculty of Physics, University of Vienna, Boltzmanngasse 5, 1090 Vienna, Austria}
\author{Dario Pisignano}
\affiliation{Dipartimento di Matematica e Fisica ``Ennio De Giorgi'', Università del Salento, via Arnesano, I-73100 Lecce, Italy}
\affiliation{National Nanotechnology Laboratory of Istituto Nanoscienze-CNR, via Arnesano, I-73100 Lecce, Italy}
\author{Daniele Gentili}
\affiliation{CNR-Istituto per le Applicazioni del Calcolo ``Mauro Picone'', Via dei Taurini 19, I-00185 Rome, Italy}
\author{Giuseppe Pontrelli}
\affiliation{CNR-Istituto per le Applicazioni del Calcolo ``Mauro Picone'', Via dei Taurini 19, I-00185 Rome, Italy}
\author{Sauro Succi}
\affiliation{CNR-Istituto per le Applicazioni del Calcolo ``Mauro Picone'', Via dei Taurini 19, I-00185 Rome, Italy}

\begin{abstract}
The effects of a driven fast-oscillating spinneret on the bending instability of electrified jets, leading
to the formation of spiral structures in electrospinning experiments with charged polymers, 
are explored by means of extensive computer simulations.
It is found that the morphology of the spirals can be placed in direct correspondence with 
the oscillation frequency and amplitude. 
In particular, by increasing the oscillation amplitude and frequency, thinner fibres can be extracted
by the same polymer material, thereby opening design scenarios in electrospinning experiments.
\end{abstract}
\maketitle

\section{Introduction}

The dynamics of charged polymer jets under the effect of an external electrostatic field stands out as major challenge in non-equilibrium thermodynamics, with numerous applications in micro and nano-engineering and life sciences as well ~\cite{Yarin,Reneker2000,Yarin2001,Loscertales2002,Dzenis2004, Helgeson2008,Thompson2007,Grafahrend2011,Greenfeld2012,Onses2013,Min2013}. Indeed, charged liquid jets may develop several types of instabilities depending on the relative strength of the various forces acting upon them, primarily electrostatic Coulomb self-repulsion, viscoelastic drag and surface tension effects. Among others, one should mention bending and ``whipping'' instabilities, the latter consisting of fast large-scale lashes, resembling the action of a whip. These instabilities are central to the manufacturing process known as electrospinning ~\cite{Regev2013,Nakano2012,Greiner2007,Fridrikh2003,Camposeo2013,Pisignano2013}. Some of them were analysed in the late sixties by G.I. Taylor \cite{Taylor1969}. In the subsequent years, it became clear that the driver of such instabilities is the Coulomb self-repulsion, as it can be inferred by simply observing that any off-axis perturbation of a collinear set of equal-sign charges would only grow under the effect of Coulomb repulsion~\cite{Reneker2000,Marin2008}.

In the electrospinning process, ultrathin fibres, with diameters in the range of hundreds of nanometres and below, are 
produced out of charged polymer jets. In addition to its fundamental importance in the fields of soft matter physics and fluid dynamics, 
electrospinning is raising a continuously increasing interest due to its widespread application fields. For this reason, nowadays this technique is
an excellent example of how applied physics impacts on engineering and materials science. 
For example, recently electrospun nanofibres and nanowires have been used for realizing organic field-effects transistors ~\cite{Liu2005, Min2013}, whose fabrication benefits from the concomitant high spatial resolution of active channels, large-area deposition, and frequently improved charge carrier mobility, which is highly relevant for nanoelectronics. Other important applications are in the field of photonics ~\cite{Camposeo2013}, and include solid-state organic lasers ~\cite{Persano2014}, light-emitting devices ~\cite{Moran2007, Vohra2011}, and active nanomaterials featuring high internal orientational order of embedded chromophores, thus exhibiting polarized infrared, Raman ~\cite{Kakade2007, Pagliara2011, Richard-L2012} and emission spectra ~\cite{Pagliara2009}. Electrohydrodynamics jet printing has been recently applied in order to control the hierarchical self-assembly of patterns in deposited block-copolymer films ~\cite{Onses2013}, with important applications in nanomanufacturing and surface engineering. Electrostatic spinning is also a tool for medical sciences and regenerative medicine, e.g. for the construction of complex scaffolding necessary for tissue growth ~\cite{Grafahrend2011, Xie2009}, or for the accurate encapsulation of solutions into monodisperse drops ~\cite{Loscertales2002}.  In general electrospinning is an excellent technique to transfer to a macroscopic composite material the properties of complex polymer solutions~\cite{Dzenis2004}, which can have a direct impact on the mechanical properties of resulting nanofibers ~\cite{Arinstein2007}. In this respect, an aspect which is particularly critical is that fibres can reach a diameter in the nanometre scale. Electrospinning can reach such scales, especially thanks to the occurrence of instabilities that drive the system to extend on the plane orthogonal to the jet axis~\cite{Yarin,Reneker2000,Yarin2001,Helgeson2008,Thompson2007,Greenfeld2012}. 
In such experiments, a droplet of charged polymer solution is injected from a nozzle at
one end of the apparatus (spinneret), then it is elongated and the resulting collinear jet moves away from the droplet under the effect of an 
externally applied electrostatic field. Such collinear configuration is however unstable against off-axis perturbations, as one can readily realize by inspecting the effect of Coulomb repulsion on different portions of the jet. The resulting bending instability often gives rise to three dimensional helicoidal structures (spirals).
Due to polymer mass conservation, the spiral structures get thinner and thinner as they
proceed downwards, until they hit a collecting plate at the bottom. 
Based on the above, it is clear that an accurate control of the effects of the bending and whipping instabilities 
on the morphological features of the resulting spirals is key to an
efficient design of the electrospinning process.
The fundamental physics of the electrospinning process is
governed by the competition between Coulomb repulsion and the stabilizing effects of viscoelastic drag and 
surface tension. Since the timespan of the entire process is comparable with the relaxation
time of the polymer material, electrospinning qualifies as a strongly off-equilibrium process.

A number of papers have been dealing with the theory of the
electrospinning process~\cite{Hohman2001,Hohman2001a,Reneker2000,Yarin,Greenfeld2011,Fridrikh2003,Pontrelli}.
Broadly speaking, these models fall within two general classes: continuum and discrete.
The former treat the polymer jet as a charged fluid, obeying the equations of
continuum mechanics, while the latter represent the jet as a discrete collection of 
charged particles (beads), subject to four type of interactions: Coulomb repulsion, viscoelastic drag, curvature-driven 
surface tension and, finally, the external electric field.

However, due to the complexity of the resulting dynamics and to the large number of experimental parameters involved in the process (related to solution, field and environmental properties), electrified jets are still treated by empirical approaches. For instance, either near-field techniques have been proposed to reduce instabilities~\cite{Sun2006,Bisht2011}, or the effect of the different parameters on the resulting nanofibre properties and radius have been determined empirically through systematic campaigns~\cite{Thompson2007,Pai2011}. Indeed, to date most investigations on electrospinning have been focussing on the experimental exploration of the various type of polymers that are liable to be electrospun into fibres, as well as on the processing/properties of the spun fibres.  

In addition, the dynamics of the jet is also sensitive to random external disturbances (noise), 
and particularly to erratic oscillations of the injection apparatus and of ambient atmosphere, which could act as hidden variables critically affecting the reliability of the process. 
In the sequel, we shall consider such fast mechanical oscillations of the spinneret as a driving perturbation, whose amplitude
and frequency can be fine-tuned in order to minimize the thickness of the electrospun fibre. 
This portrays an angle of investigation of the electrospinning 
process that we proceed to investigate, on a systematic basis in this paper. In particular, it is found that either increasing the perturbation amplitude, frequency, or both can lead to obtain thinner jets, hence thinner fibres in the end. More precisely, by increasing the
perturbation amplitude, spirals are seen to open up into a broader cone envelope, hence resulting into
thinner fibres. By increasing the perturbation frequency, on the other hand, the spirals are developed with a shorter pitch, resulting again in thinner fibres at the collecting plate.
To highlight the potential of driving the injected fluid to produce thinner fibres in practice, we point out that increasing the perturbation amplitudes, from say $1.6\;10^{-4}\;\text{cm}$ to $1.6\;10^{-3}\;\text{cm}$, while keeping the other process parameters unchanged, leads to a three-fold reduction of the resulting fibre thickness for polyethylene
oxide and other plastic materials. Similarly, varying the perturbation frequency from $10^{5}\;\text{s}^{-1}$ to $10^{6}\;\text{s}^{-1}$ determines a three-fold fibre thickness decrease.

\section {Mathematical model}

The mathematical model used in this paper closely follows the one given in the pioneering work by Reneker and coworkers ~\cite{Reneker2000}, namely the jet is described by a sequence of discrete charged particles (beads), obeying Maxwell fluid mechanics under the effects of the four forces described above.
The polymer jet is represented by means of a Maxwellian liquid, coarse-grained 
into a viscoelastic bead-spring model with a charge associated to each bead. 
This representation is justified by the earlier work of Yarin~\cite{Yarin1990,Yarin}.
In Fig. 1 we show how the experimental set-up looks like. On the top of the figure there is the pendant drop, which is dangling from a pipette or from a syringe. Between the pipette placed along the $Z$-axis at a height $h$ and the collecting plane (not visible in the figure) at $Z=0$, the electric field $V_0/h$ is applied downwards along the $Z$ direction. 
\begin{figure}[ht]
\begin{center}
\includegraphics[width=1.0\columnwidth]{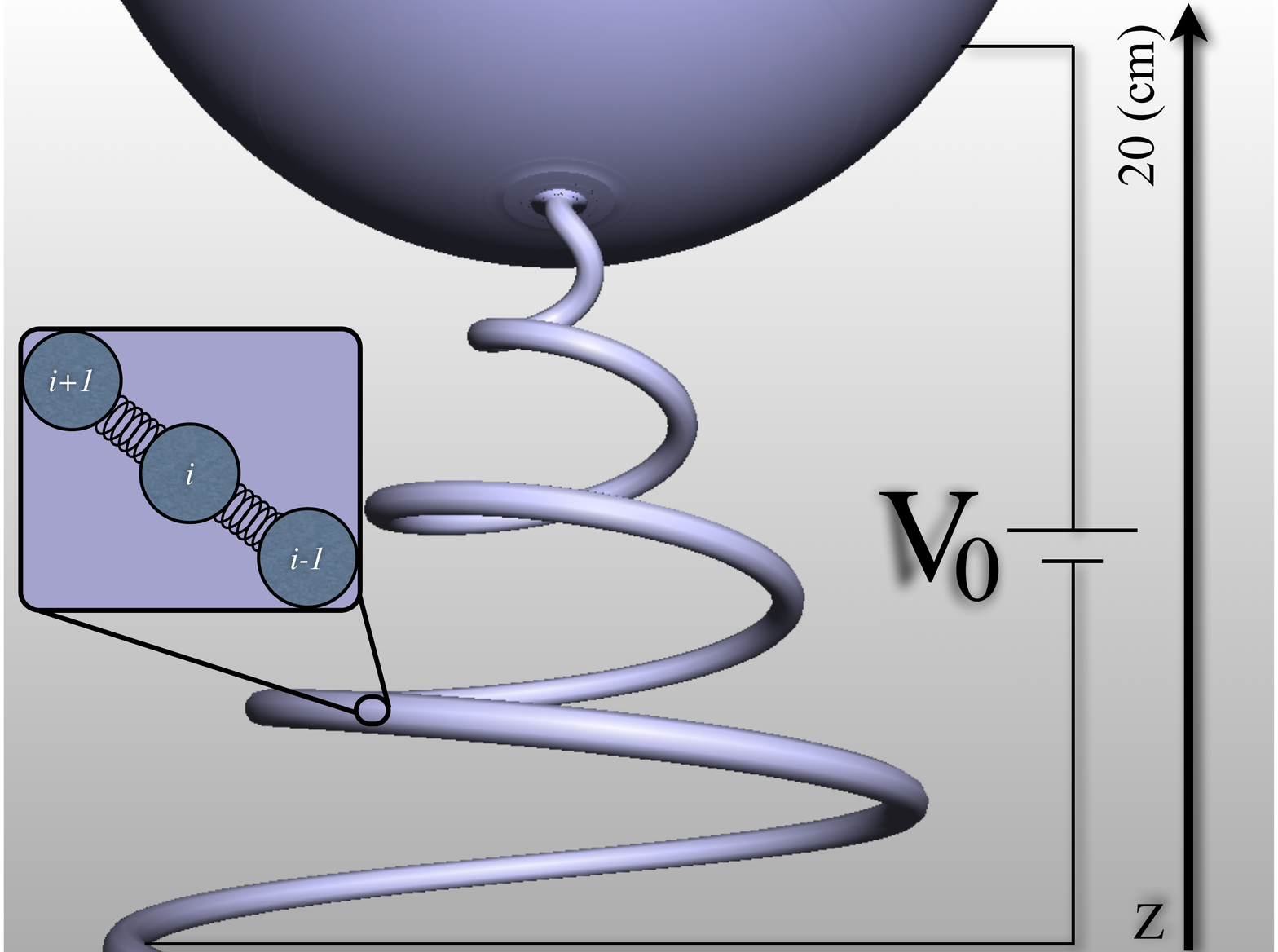}
\caption{Schematic representation of the model for the polymer jet. The jet is generated from the pendant drop on the top of the figure upon application of the external field $V_0/h$. The jet is then represented at a microscopic level as a bead-spring chain. Each bead is charged while the connecting springs are calculated by solving the associated  polymer stress equations at each molecular dynamics time step.}\label{model-sketch}
\end{center}
\end{figure}
We shall consider the following experimental parameters: $a_0$ is the initial cross-section radius of the electrified jet, $e$ is the charge per particle, $G$ is the elastic modulus, $h$ is the distance from the drop to the collector, $m$ is the mass of each particle, $\mu$ is the dynamic viscosity, $\theta=\mu/G$ is the relaxation time, $\alpha$ is the surface tension, and $V_0$ is the voltage applied between the drop  and the grounded collector plate. In what follows we will use Gaussian units for charges (CGS units).
In Fig. 1, the spring represents the stress $\sigma$ in the Maxwellian fluid and it is determined by the stress equation between each consecutive pairs of beads. The coupled system of Newton equations for the beads and the associated stress equation reads as follow:
\begin{equation}
\begin{cases}
&m\frac{d^2\vec{R_i}}{dt^2}= \vec{F_i} \\
&\frac{d\sigma_i}{dt} =  \frac{1}{l_i}\frac{dl_i}{dt} -\sigma_i, 
\end{cases}\label{stress-equation}
\end{equation}
where $\vec{R_i}=\left(X_i,Y_i,Z_i\right)$ is the position vector of the i-th bead, $l_i=\left[{\Delta X_i}^2+{\Delta Y_i}^2+{\Delta Z_i}^2\right]^{1/2}$ is the distance between two bonded beads. Space and time units have been chosen as  $t=t_{\text{phys}}/\theta$ and $l=l_{\text{phys}}/L$, where $L=(e^2/\pi a^2_0 G)^{1/2}$. In the sequel, the superscript (subscript) U (respectively D), will refer to the ``up'' $(i+1)$ and ``down''  $(i-1)$ beads, respectively, where $i$ is the bead index. The coupling between the equations Eq.~\ref{stress-equation} takes place through  the net viscoelastic force $ \vec{F}_{\texttt{VE}}$ acting on the bead $i$, which is given by:

\begin{eqnarray}
  \vec{F}_{\texttt{VE}}=F_{ve}&&\left [\left (a^2_U\sigma^U_i\frac{X_{i+1}-X_i}{l^U_i}-a^2_D\sigma^D_i\frac{X_i-X_{i-1}}{l^D_i}\right)\vec{i}\right. \nonumber \\
&& +\left(a^2_U\sigma^U_i\frac{Y_{i+1}-Y_i}{l^U_i}-a^2_D\sigma^D_i\frac{Y_i-Y_{i-1}}{l^D_i}\right)\vec{j} \nonumber \\
&& \left. +\left(a^2_U\sigma^U_i\frac{Z_{i+1}-Z_i}{l^U_i}-a^2_D\sigma^D_i\frac{Z_i-Z_{i-1}}{l^D_i}\right)\vec{k}\right]\label{F-visco-elastic}
  \end{eqnarray}
where in reduced units one has $a^2_U=1/l^U_i$ and $a^2_D=1/l^D_i$.
Besides the viscoelastic force $ \vec{F}_{\texttt{VE}}$, each bead is subject to three additional forces, namely: the driving force $\vec{F_0}$ of the external field $\vec{V_0}$, acting between the drop and the collection plate, the bead-bead Coulomb interaction $\vec{F}_{\texttt{Coul}}$, and the surface tension force $\vec{F}_{\texttt{Cap}}$, which acts as an effective bending rigidity penalising curved jet shapes. These forces read as follows:

\begin{eqnarray}
\vec{F_0}&=&-\frac{e\mu^2 |\vec{V_0}|}{h L m G^2}\vec{k}\nonumber \\
\vec{F}_{\texttt{Coul}}&=&Q\sum_j\left(\frac{X_i-X_j}{R_{ij}^3}\;\vec{i}+\frac{Y_i-Y_j}{R_{ij}^3}\;\vec{j}+\frac{Z_i-zß_j}{R_{ij}^3}\;\vec{k}\right) \nonumber \\
\vec{F}_{\texttt{Cap}}&=&A\;k_i\left(-X_i\frac{(a_U+a_D)^2}{4 \sqrt{X^2_i+Y^2_i}}\;\vec{i}-Y_i\frac{(a_U+a_D)^2 }{4 \sqrt{X^2_i+Y^2_i}}\;\vec{j}\right) \label{forces}
\end{eqnarray}
where $F_{ve}\equiv Q=\frac{\pi a^2_0 \mu^2}{LmG}$, $A=\frac{\pi\alpha a^2_0 \mu^2}{m L^2 G^2}$, $R_{ij}$ is the distance between two beads, and $k_i$ is the local curvature, defined as the radius of the circle going through the points $i+1$, $i$ and $i-1$. 
It is important to stress that the model also assumes that the fluid is incompressible, which results in the mass conservation law:
\begin{equation}
\pi a^2 l = \pi a_0^2 l_0
\end{equation}
This determines the fibre thickness $a$ as a function of the elongation $l$.
At each time step, we first integrate the above stress equations Eq.~\ref{stress-equation}, and then we use the updated stress terms to integrate the equations of motion which, in turn, are solved with a simple velocity Verlet integration scheme with the same integration time step $\Delta t$ used for the stress equation. 

\section{Numerical integration scheme}

We solve  Eq.~\ref{stress-equation} in an equivalent but numerically more convenient form:
\begin{equation}
\frac{d\left(e^t\sigma\right)}{dt}=e^t\left(\frac{1}{l}\frac{dl}{dt} \right)
\end{equation}
using the forward Euler discrete integration scheme with an integration step $\Delta t$:
\begin{equation}
e^{t}\;\sigma(t) - e^{t-\Delta t}\;\sigma(t-\Delta t)=\Delta t\;e^{t}\left(\frac{1}{l(t)}\frac{l(t)-l(t-\Delta t)}{\Delta t} \right)
\end{equation}
which leads to the explicit time marching scheme
\begin{equation}
\sigma(t) = e^{-\Delta t}\;\sigma(t-\Delta t)+ \frac{l(t)-l(t-\Delta t)}{l(t)}\label{Euler_Stress}
\end{equation}

The Euler integration is coupled to a Verlet time integration scheme as follows, given the initial conditions at $t=-\Delta t$ and at $t=0$ :
\begin{enumerate}
\item  Stress at time $t$: $\sigma(t-\Delta t) \rightarrow \sigma(t)$ from Eq.~\ref{Euler_Stress};
\item  Forces at time $t$:  $\vec{F}(t)=\vec{F}_{\texttt{VE}}+\vec{F}_{\texttt{Coul}}+\vec{F}_{\texttt{Cap}}+\vec{F_0}$;
\item  Positions at time $t+ \Delta t$: $\vec{r}(t+ \Delta t)=2\vec{r}(t)-\vec{r}(t-\Delta t)+\vec{F}(t) \Delta t^2$;
\item Velocities at time $t+ \Delta t$: $\vec{v}(t+ \Delta t)=(\vec{r}(t+ \Delta t)-\vec{r}(t - \Delta t))/(2\Delta t)$.
\end{enumerate}  
We wish to point out that the $1/r^2$ Coulomb singularity needs to be handled
with great care, especially at the injection stage, where beads are 
inserted at a mutual distance shorter than their linear size.
This imposes very small time-steps, of the order of $10^{-8} \mu/G$.

The effect of mechanical oscillations is mimicked by injecting the tail 
bead $i=N(t);\vec{V}^i_{\text{Init}}=0$, $N(t)$ being the number of beads in the chain at time $t$, at 
an off-axis position given by:
\begin{eqnarray}
X_N = N_s L \cos(\Omega t) \label{LOADx}\\
Y_N = N_s L \sin(\Omega t) \label{LOADy}\\
Z_N = h-L_{ins}
\end{eqnarray}
where $L_{ins}=h/I_F$ is the insertion length and $I_F$ the insertion factor, $h$ the vertical size of the apparatus, typically $I_F=50000$, $N_s \ll 1$ is the amplitude and $\Omega$ is the frequency of the dynamical perturbation.
In the sequel we shall refer to $N_s$ as to the noise strength.
We remind that the insertion algorithm proceeds by inserting the $N$-th bead (polymer jet tail) once the
distance between the $(N-1)$-th bead and the pendant drop at $Z=h$ exceeds the 
insertion distance, $L_{ins}=h/I_F$. 
Thus, the jet is represented by a bead chain with the tail, $i=N$, at the spinneret and the head, $i=1$, 
proceeding downwards to the collector. The idea behind this model~\cite{Reneker2000} is that, by choosing $\Omega \gg 1$, the above algorithm would generate a quasi-random sequence of initial slopes of the polymer bead chain, whose envelope defines a conical surface known as the Taylor cone. In this work, however, the expression (\ref{LOADx}-\ref{LOADy}) stands for a deterministic, {\it controlled} source of fast oscillations, whose amplitude and frequencies can be fine-tuned to design thinner fibres. To quantitatively explore this idea, we have run simulations 
with several values of the perturbation strength $N_s$ and perturbation frequency $\Omega$. 

Boundary conditions are imposed at the two ends of the jet: at $Z=h$ (top) and 
at the collecting plate $Z=0$ (bottom).
The latter is treated as an impenetrable plane, at which the Coulomb forces are set to zero and
the $Z$ component of the bead position is not allowed to take negative values.
The top boundary condition is such that the $X$ and $Y$ components of the forces 
acting on the pendant drop are set to zero, while the $Z$ component must be non positive, i.e. point downwards.

In what follows, we will use the scheme above to solve the time development of the jet with a reference set of parameters pretty close, yet not exactly equal to the one given in
the original paper by Reneker et al.~\cite{Reneker2000} as presented in Table~\ref{tab:Real_Values}. For the case in point $L=0.32$ cm and $\theta=0.01$ s. The resulting dimensionless parameters are given in the right panel of Table I, where $Q$, $F_{ve}$ and $A$, are the strength of Coulomb repulsion, viscoelastic forces and surface tension, respectively, $K_s=\Omega\;\theta$ is the injection frequency, $H=h/L$ and $\vec{F_0}$ is the scaled force resulting form the external field, all in dimensionless units.

\newsavebox\Realvalues
\begin{lrbox}{\Realvalues}
  \begin{minipage}{0.4\textwidth}
    \begin{eqnarray}
\alpha &=& 700 \, \text{g}/\text{s}^2\nonumber \\ 
a_0 &=& 1.5\;10^{-2} \, \text{cm}\nonumber \\ 
e &=& 8.48 \, \frac{g^{1/2} \text{cm}^{3/2}}{\text{s}}\nonumber \\ 
\theta &=& 0.01 \, \text{s}\nonumber \\ 
G &=& 10^{6} \, \frac{\text{g}}{\text{cm}\;\text{s}^2}\nonumber \\ 
h &=& 20\;\text{cm}\nonumber \\ 
m &=& 2.83\;10^{-6}\;\text{g}\nonumber \\ 
\mu &=& 10^{4} \;\frac{\text{g}}{cm\;s}\nonumber \\ 
V_0 &=& 10^{4} \;\text{V}\nonumber \\ 
\Omega &=& 10^{4} \;\text{s}^{-1}\nonumber
\end{eqnarray}
  \end{minipage}
\end{lrbox}

\newsavebox\Param
\begin{lrbox}{\Param}
  \begin{minipage}{0.3\textwidth}
    \begin{eqnarray}
Q &\equiv & F_{ve} = 78309.8\;\text{(Eq.\ref{forces})}\nonumber \\ 
F_0 &=& 157.9 \;\text{(Eq.\ref{forces})}\nonumber \\ 
A &=& 171.9 \;\text{(Eq.\ref{forces})}\nonumber \\ 
K_s &=& 100\nonumber \\
H &=& 62.8\nonumber  
\end{eqnarray}
  \end{minipage}
\end{lrbox}

\begin{table}
  
  \caption{Simulation parameters obtained from the experimental data in the paper of Reneker et al.~\cite{Reneker2000}}
  \begin{center}
    \begin{tabular}{cc}
       Experimental Parameters & Simulation Parameters \\ \hline \hline
       \usebox{\Realvalues} & \usebox{\Param} \\
       \label{tab:Real_Values}
    \end{tabular}
  \end{center}
\end{table}

\section{Results and Discussion}
  
\begin{figure}[ht]
\subfigure[\label{Figure1}]{\includegraphics[width=0.49\columnwidth]{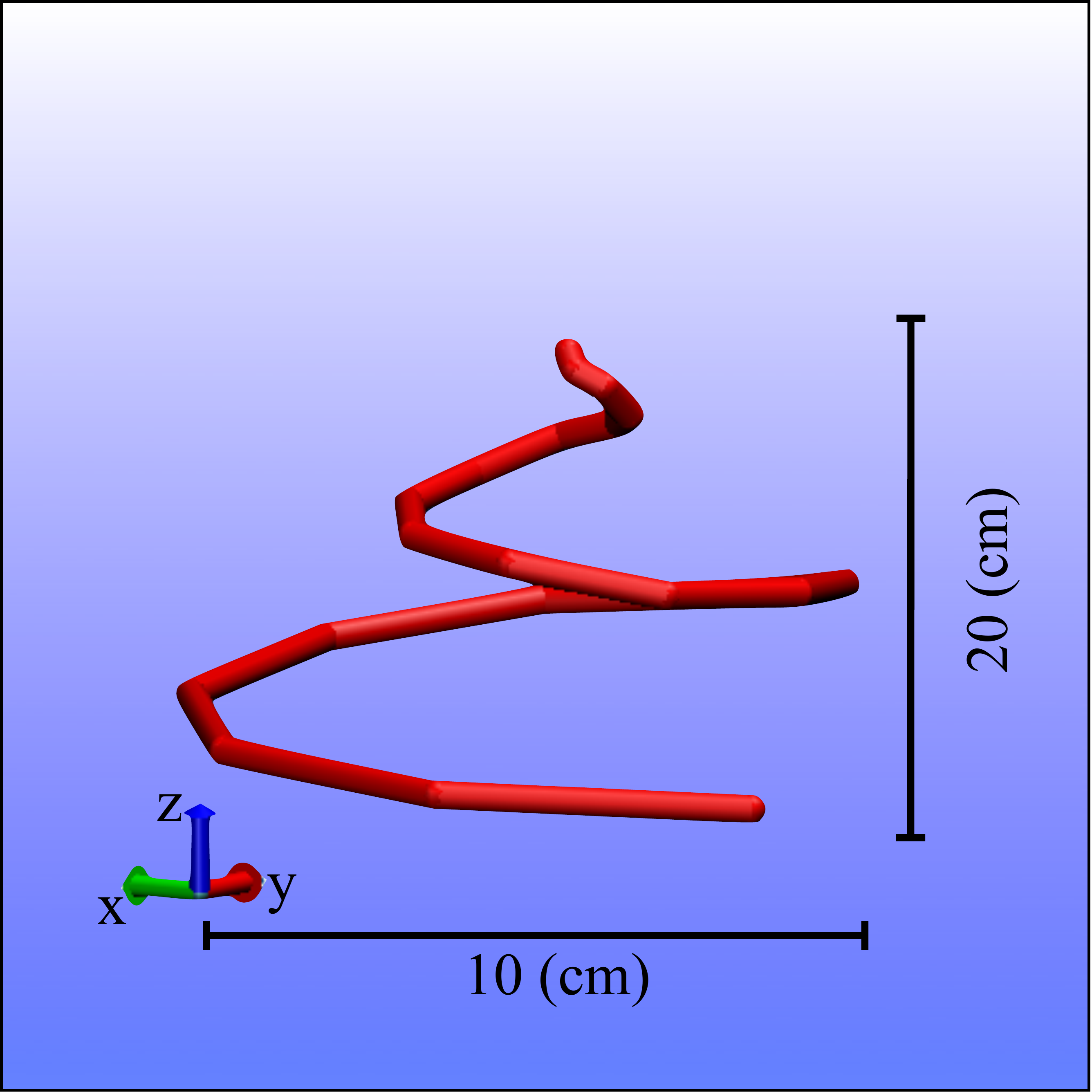}} 
\subfigure[\label{1D-conf}]{\includegraphics[width=0.49\columnwidth]{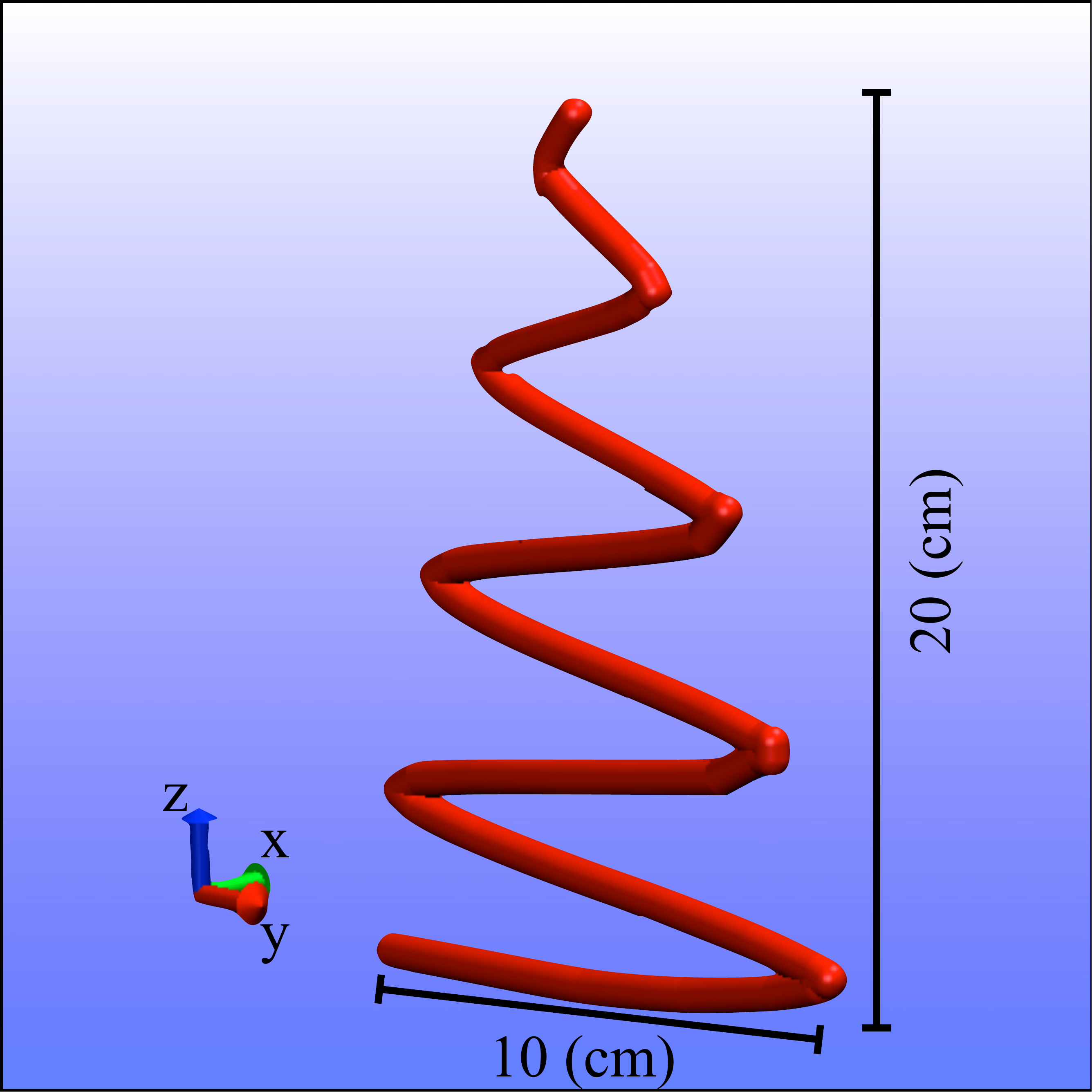}}
\caption{Final configuration of the filament in the case of a circular driving perturbation (a) and a linear one (b). Parameters used for the simulations are $ V_0=10^{4}$ V, $\Omega=5 \;10^5\;\text{s}^{-1}$, $N_s=10^{-3}$ and the total spinneret-collector distance $ h=20$ cm, which resulted in the rescaled quantities $ L= 0.32 $ $Q=F_{ve}= 78309.8$, $ F_0= 157.9$, $ A= 171.9$, $ H= 62.8$. The linear dimensions are depicted in the figure and correspond to the following range of the axes: $X\in[-5,5]$ cm, $Y\in[-5,5]$ cm and $Z\in[0,20]$ cm.} 
\end{figure}

In Fig.~\ref{Figure1}, we show a typical shape of the jet at the 
end of a simulation, lasting about $10^6$ time steps.
As one can appreciate, spirals show from three to about ten turns over distances of a few centimetres, which is consistent with experimental observation~\cite{Reneker2008,Bhattacharjee2011}.
While we cannot rule out the possibility that kinematic effects play a major role in the spiral formation, we must observe that the morphology of the spiral appears highly dependent on the choice of the initial conditions and forcing parameters. In this respect, even the kinematics alone  may not be trivial at all.
We also considered a scenario whereby the perturbation is not applied on the $XY$-plane, but only along the $X$-axis.
Since the component of the force along the $Y$-axis is always zero, the jet does not develop 
any component in the $YZ$-plane, resulting in a flat sinusoidal profile in the $XZ$-plane (Fig.~\ref{1D-conf}). Also these classes of planar spirals are frequently observed in experiments, and not explored in previous theoretical papers. 
The extreme variability of experimentally observed spirals can therefore be accounted for by considering 
the oscillating perturbation as an independent variable affecting the resulting dynamics of the electrified jets.
  

\subsection{Effect of the perturbation amplitude}

\begin{figure}[ht]
\subfigure[]{\includegraphics[width=0.70\textwidth]{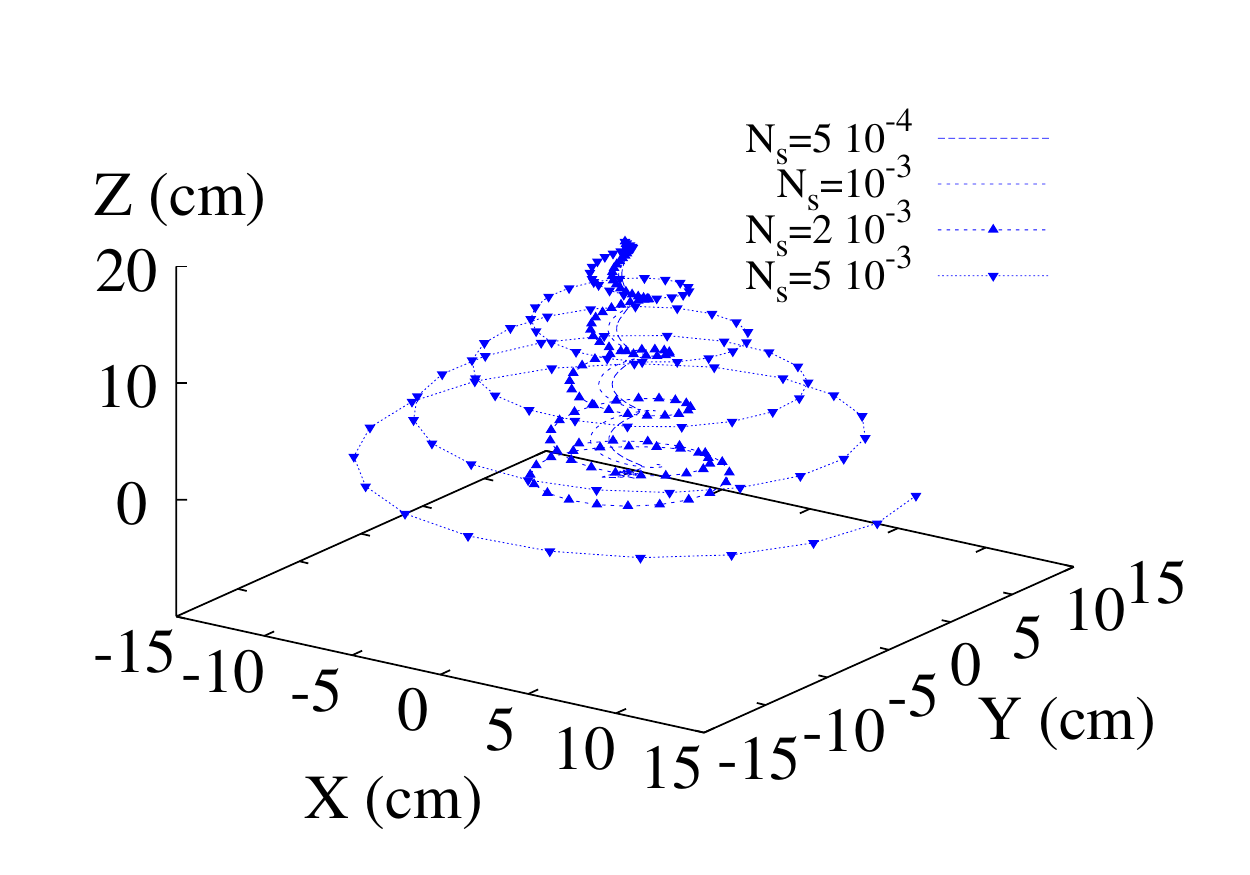}}
\subfigure[]{\includegraphics[width=0.70\textwidth]{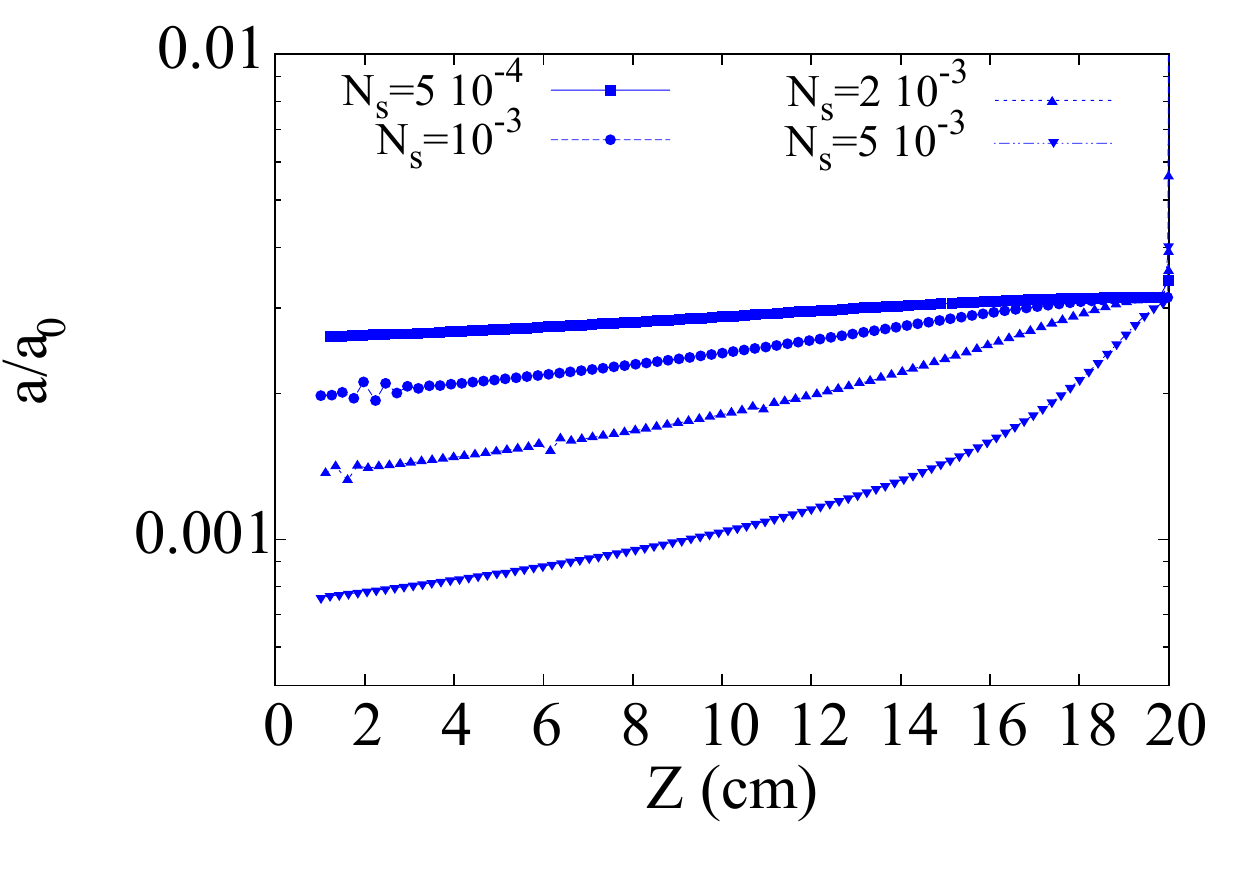}}
\caption{Effect on the spiral shape of the mechanical perturbation applied to the pendant drop. 
(a) Final configuration. Parameters used for the simulations are as above except for $N_s$. The plot shows the comparison of the profiles obtained with different 
values of the perturbation strength $N_s$. (b) Plot of the elongation along the spiral as a function of the distance from the collector. The plot shows a monotonic decrease of the fibre thickness as a function of the distance from the pendant drop. Finally, the plot highlights that an increase in the spiral radius induces an increase in the elongation of the electrified jet and correspondingly a reduction of the diameter of the resulting fibres.
}\label{Figure-2}
\end{figure}

One of the environmental parameters of the model is the amplitude of the noise 
on the pendant drop, related to the vibrations that affect the experiment at the microscale. 
In the paper of Reneker et al.~\cite{Reneker2000}, the noise amplitude was kept fixed to $N_s=10^{-3}$. In Fig.~\ref{Figure-2} we show the comparison between the results
in the amplitude range $N_s\in \left[5\;10^{-4},\ldots,5 \;10^{-3}\right]$. The main effect of increasing the amplitude is to increase the aperture 
of the spiral, basically in linear proportion. This indicates that the spiral keeps full memory of its initial slope, which is consistent
with the deterministic nature of the model. 

In  Fig.~\ref{Figure-2}(b), we show the running draw ratio $a(Z)/a_0$ along the 
polymer chain, consisting of about hundred discrete beads, for
different values of the noise strength $N_s$.
The final draw ratio is computed according to the standard expression resulting from mass
conservation between the head (index 0) and tail (index $f$) dimers in the chain, namely 
$\frac{a_f}{a_0}= \sqrt{\frac{cL_{\text{ins}}}{l_{\text{final}}}}$, where $c=0.06$ is the initial polymer concentration
in the solution and $l_{\text{final}}$ is the elongation of the bead closer to the collecting plane at end of the simulation. As expected by mass conservation, at all values of $N_s$, the jet thickness is a decreasing 
function of the distance from the spinneret ($Z=20$).
Such a decreasing trend is more and more pronounced as $N_s$ is increased.
For the bead closest to the collector, the draw ratio goes from 
$a_f/a_0=2.5 \;10^{-3}$ for $N_s = 5\;10^{-4}$ to about $a_f/a_0=7.5 \;10^{-4}$ for $N_s=5 \;10^{-3}$.
Thus, an order of magnitude increase in the perturbation amplitude leads to
a factor three reduction in the fibre thickness. 

\begin{figure}[p]
\subfigure[]{\includegraphics[width=0.70\columnwidth]{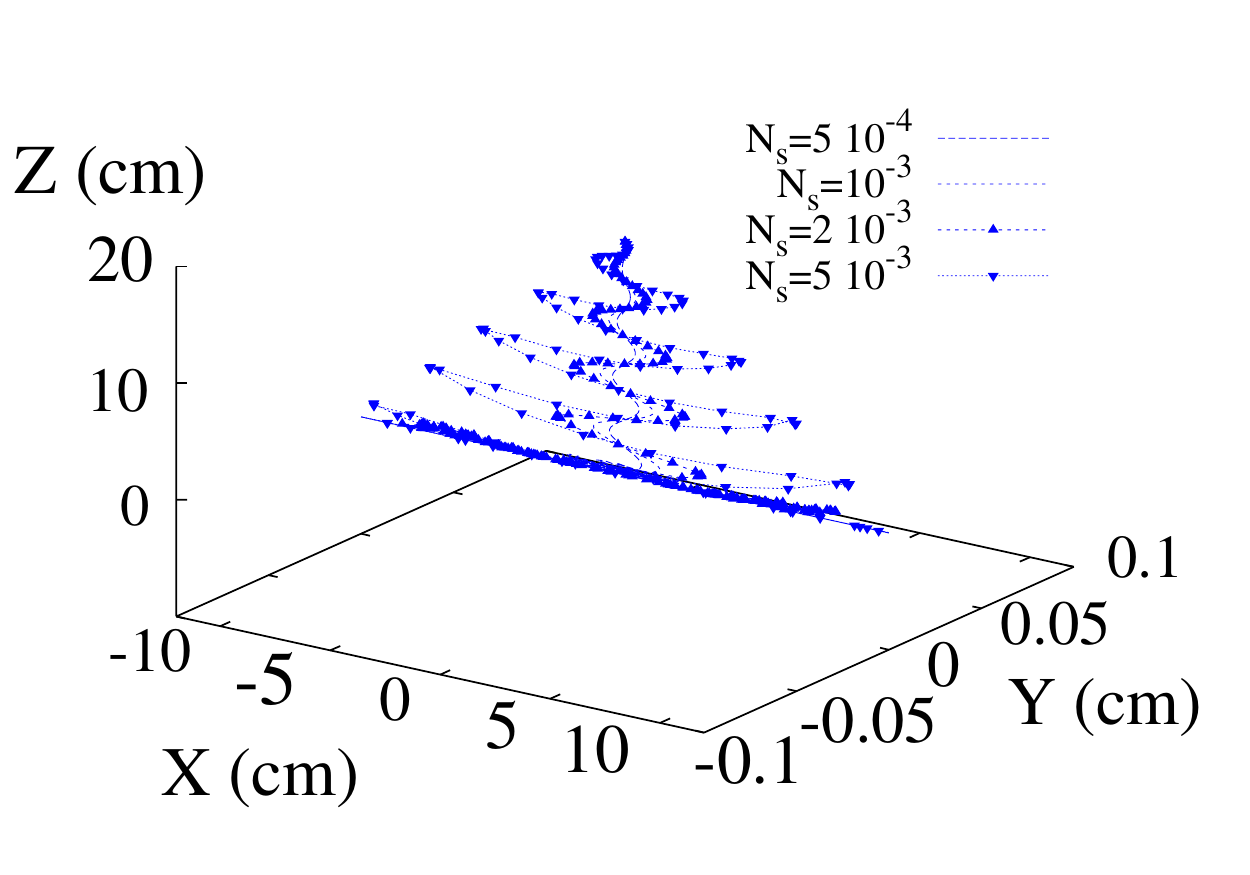}}
\subfigure[]{\includegraphics[width=0.70\columnwidth]{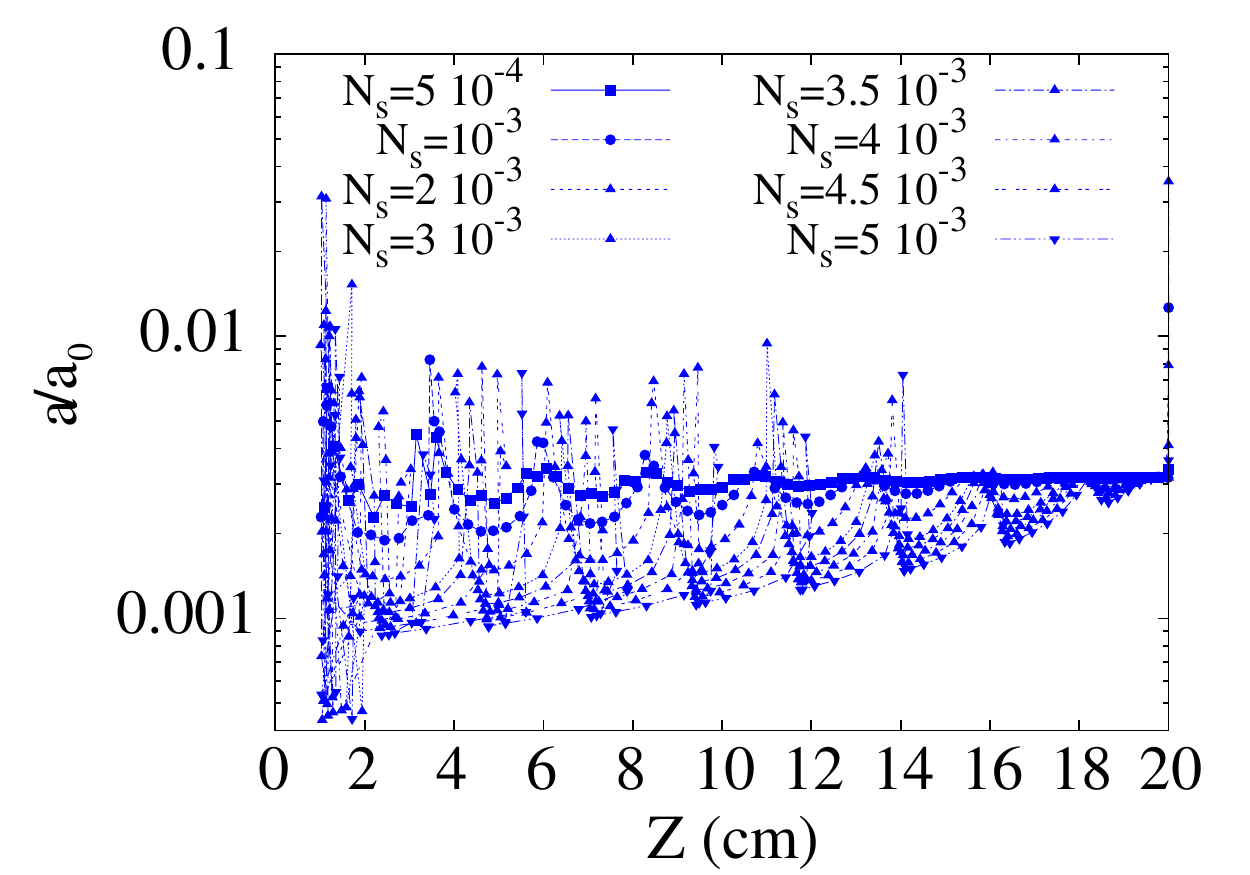}}
\caption{Effect on the spiral shape of the perturbation applied to the pendant drop only along the $X$-axis. a) Final configuration of the filament. Parameters used for the simulations are $V_0=\;10^4$ V, $\Omega=5 \;10^5 \;\text{s}^{-1}$ and $ h=20$ cm. The plot compares the profiles obtained with different 
values of the noise strength $N_s$, with the jet oscillating only in the $XZ$-plane. (b) Plot of the elongation as a function of the distance from the collector. The spikes, not present in the case of the circular applied perturbation, correspond to the bending points that occur along the $Z$-axis.
}\label{1D-noise-amplitude}
\end{figure}

By inspecting the extension of the jet, it is seen that, on average, the thickness is still reduced  by a similar
amount as in the case of the planar-perturbation (Fig.~\ref{1D-noise-amplitude}). 
However, at variance with the case of planar-perturbation, the elongation shows recurrent oscillations, due to the fact
that, close to the turning points, the fibre gets compressed, resulting in a local increase of its thickness.
Such a compression, which is visible in Fig.~\ref{1D-conf} through the blobs at the turning points, is
a purely two-dimensional effect, since in three-dimensions the bead can turn 
back by taking a {\it smooth} round trip around the $Z$-axis.  This two-dimensional topological constraint leads to the peculiar banana-like shape 
of the oscillation at high values of the perturbation amplitude, $N_s=5 \;10^{-3}$ 
(Fig.~\ref{1D-noise-amplitude}(a)), which results in large  spikes of the fibre thickness (Fig.~\ref{1D-noise-amplitude}(b)).
 Such spikes are clearly undesirable in an experimental setting, since they would introduce a critical 
dependence of the fibre thickness on the location of the collecting plane. Depending on the height of the plane, or fluctuations of the jet,
the fibre accumulation point on the collection plane would correspond either to a maximum or a minimum thickness. Moreover, one should also take into account
the drying process of the fibre that could freeze the spikes into the final fibre.
Thus, even though the qualitative morphological differences induced by dimensionality do
not significantly affect the {\it time-averaged} behaviour of the fibre diameter, compared to a pure horizontal driving  force, the planar perturbation is not only more realistic, but also definitely more convenient from the experimental viewpoint.

\begin{figure}[ht]
\subfigure[\label{FigExtension-NY}]{\includegraphics[width=0.49\columnwidth]{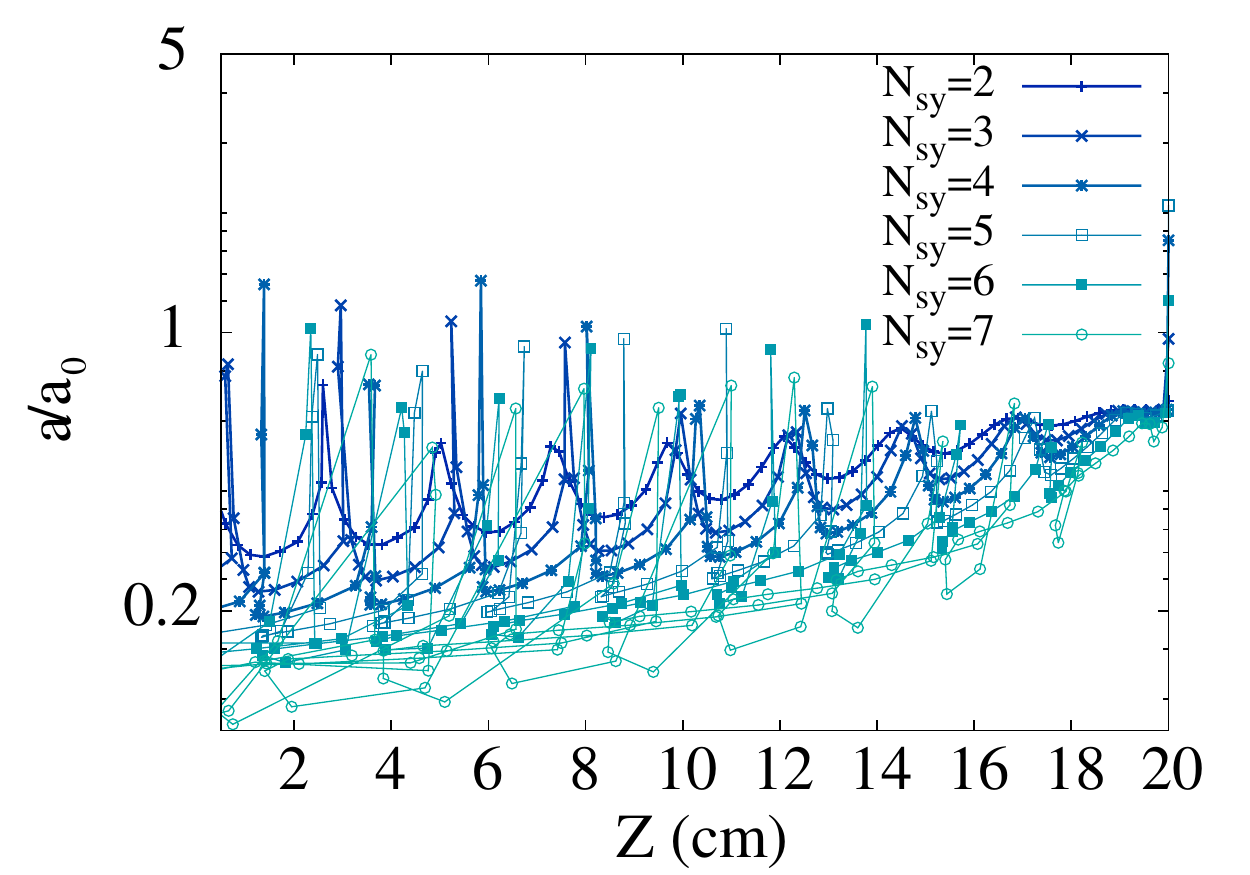}}
\subfigure[\label{FigExtension-KSZ}]{\includegraphics[width=0.49\columnwidth]{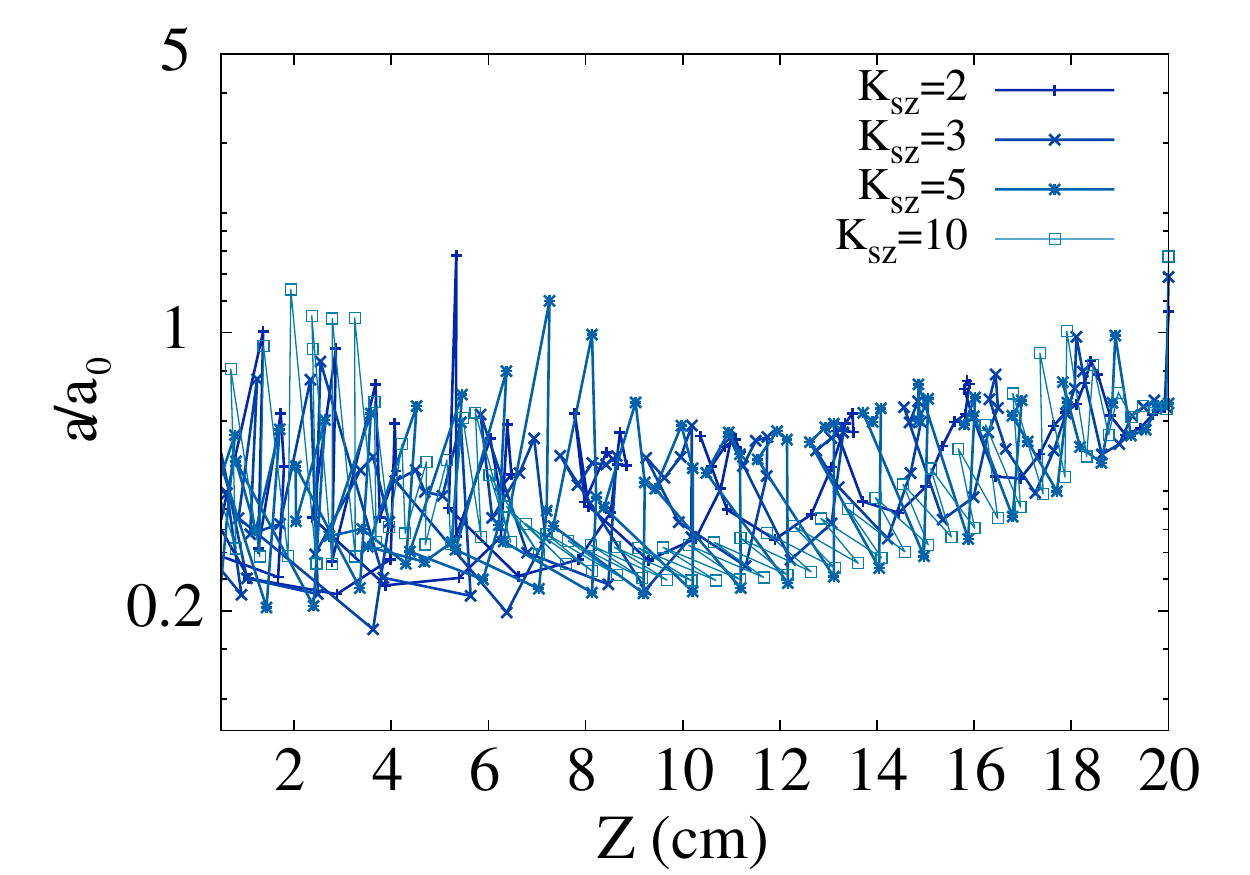}}
\caption{Plot of the fibre thickness $a/a_0$ as a function of the height from the the collecting plane at $Z=0$. We considered both asymmetric planar noises ($X/Y$ noise ratio $N_{sy}>1$) and vertical noises along the $Z$-axis ($XY/Z$ relative noise amplitude $N_{sz}=1$ and relative frequency $K_{sz}>1$). Parameters used for the simulations are $ V_0=\;10^4 $ V, $I_F=5 \;10^4$, $N_s=10^{-3}$, $h=20$ cm. As for the 1D noise scenario the elongation profile strongly fluctuates with the height from the collecting plate, making asymmetric or vertical driving forces not desirable for optimal experimental set up.}
\label{FigExtension-N}
\end{figure}

\begin{figure}[ht]
\subfigure[\label{FigNoiseFre-KSZ}]{\includegraphics[width=0.49\columnwidth]{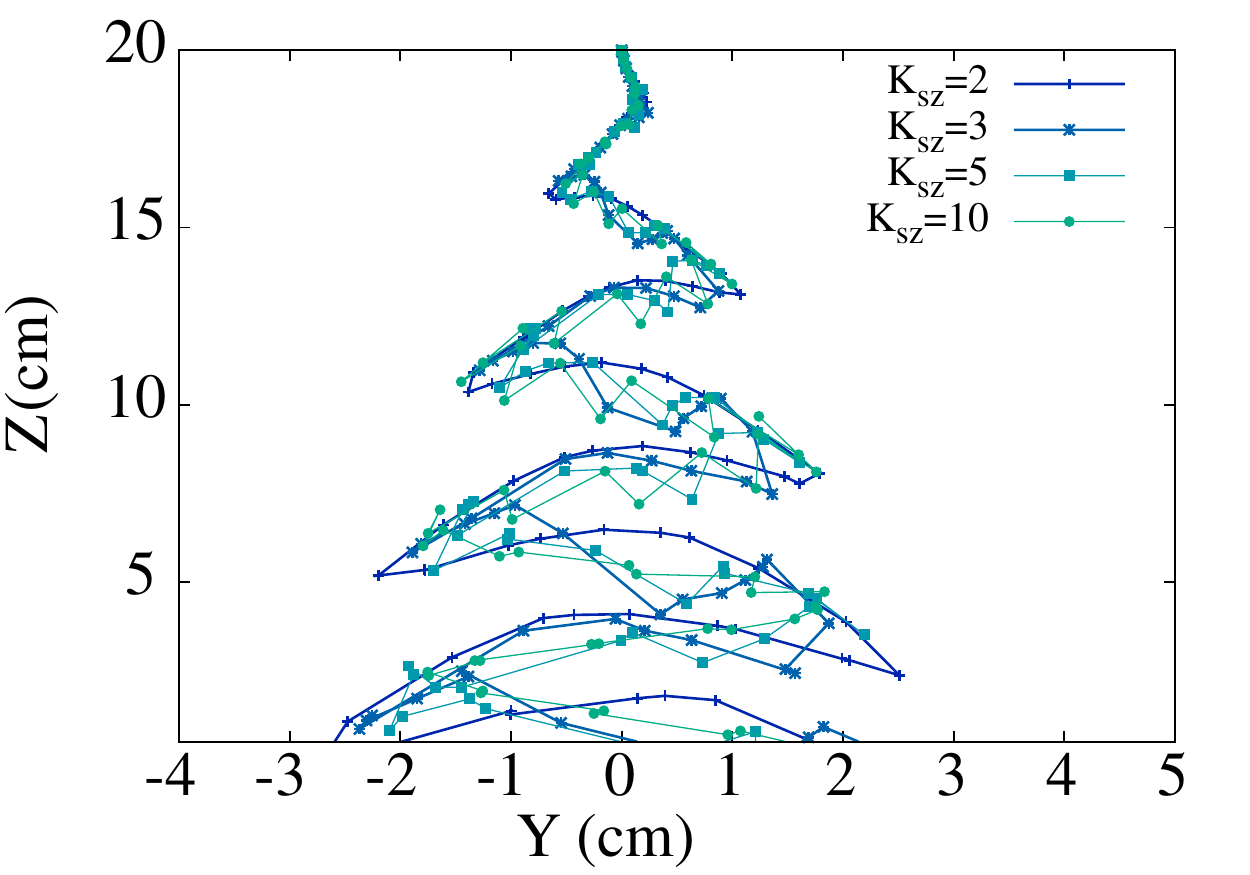}}
\subfigure[\label{FigNoiseFre-NZ}]{\includegraphics[width=0.49\columnwidth]{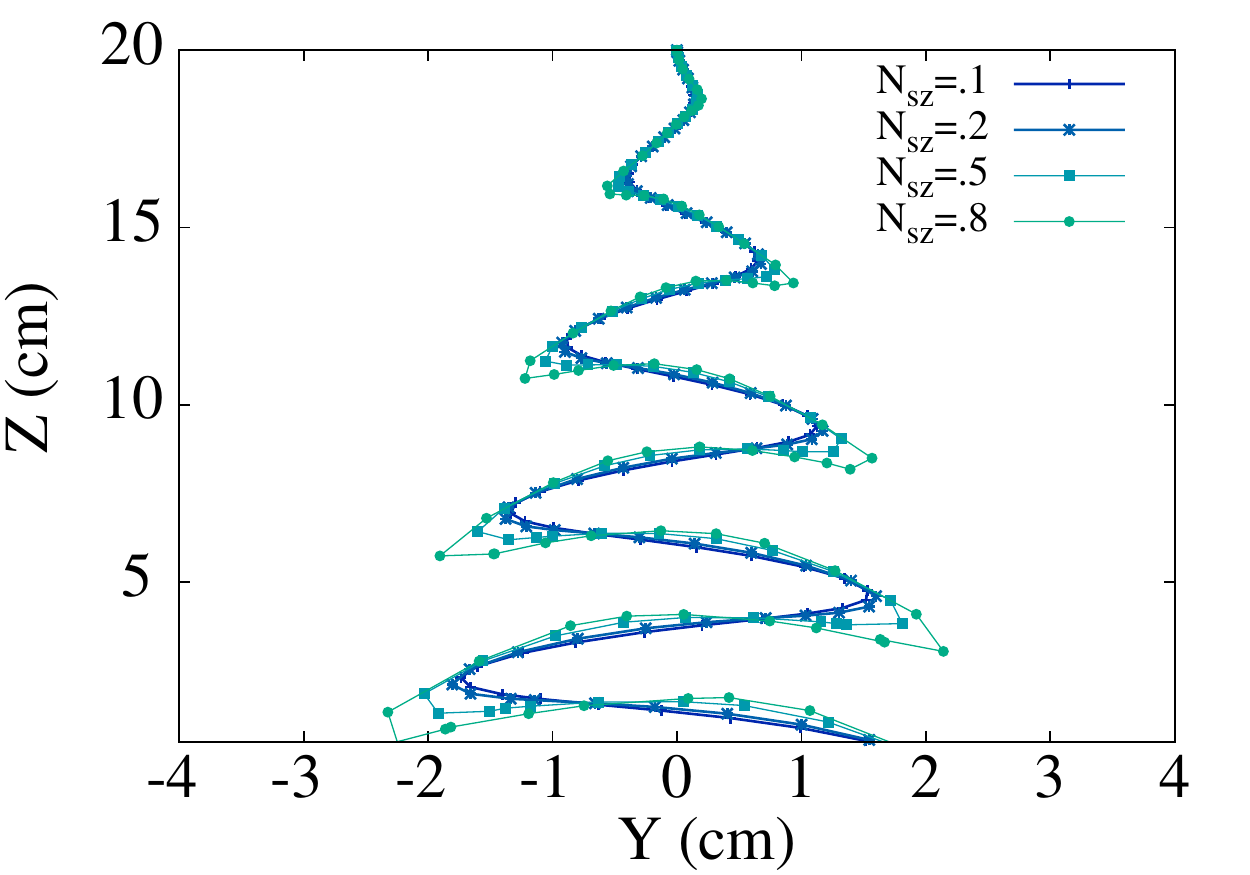}}
\caption{Spiral jet generated under the presence of a vertical driving force along the $Z$-axis on the pendant drop with a symmetric planar noise ($N_{sy}=1$). For clarity we plotted only the projection on the $ZY$-plane, but it should be noted that the spirals all keep they circular symmetry on the $XY$-plane. The parameters used for the simulations are $ V_0=\;10^4 $V, $ I_F=5 \;10^4 $, $ h=20$ cm. (a) Spiral structures for different values of the relative load frequency $K_{sz}$ to the planar noise frequency $K_s=5\;10^3$, and a fixed noise strength $N_{sz}=N_{s}=10^{-3}$. 
(b) Jet profile plotted for a range of relative noise intensities $N_{sz}$ with respect to the $XY$-plane noise $N_{s}=10^{-3}$ at a fixed frequency $K_s=5\;10^3$. The results indicate that even large vertical noises do not affect considerably the shape of the helix and in particular they do not lead to thinner fibres.}
\label{FigNoiseFre-N}
\end{figure}

\begin{figure}[ht]
\subfigure[\label{FigNoiseFre-NY-side}]{\includegraphics[width=0.49\columnwidth]{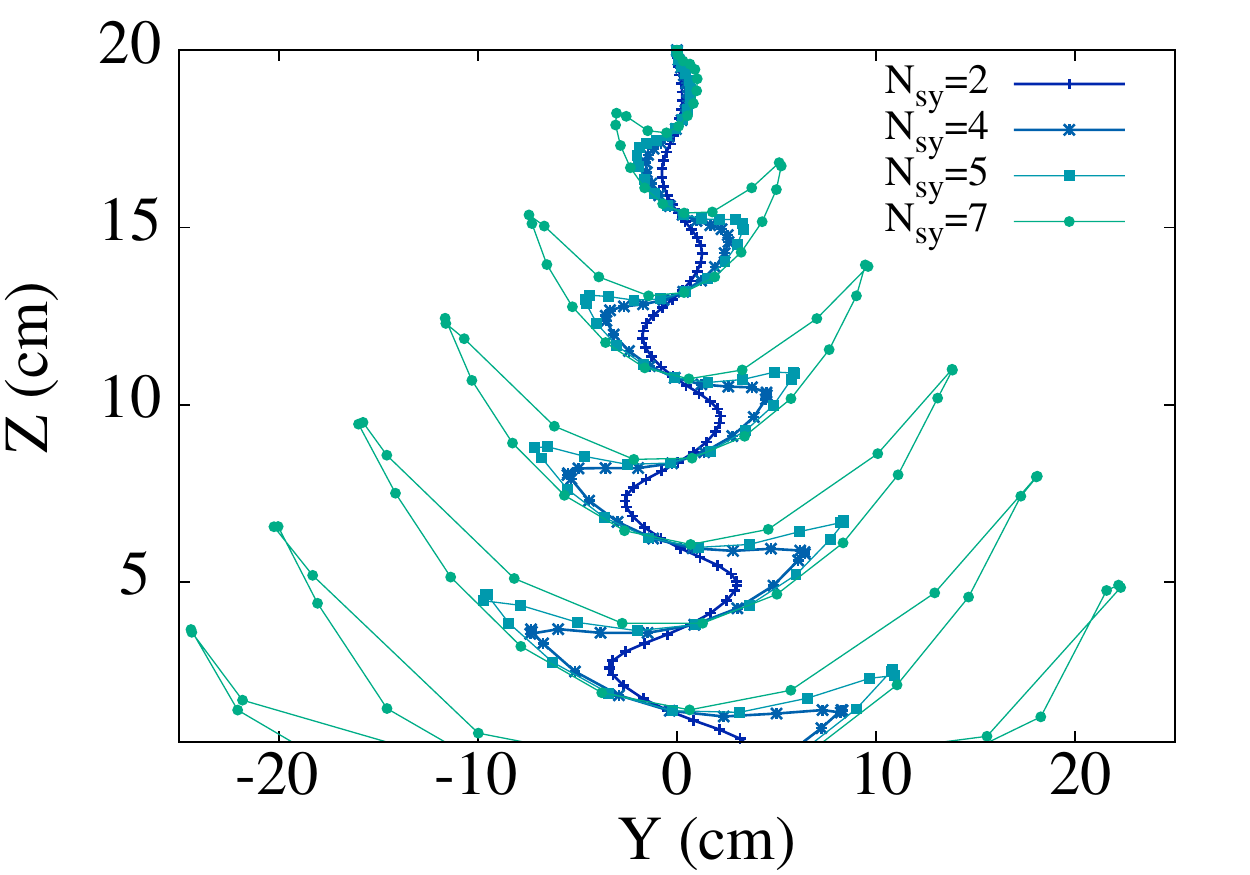}}
\subfigure[\label{FigNoiseFre-NY-top}]{\includegraphics[width=0.49\columnwidth]{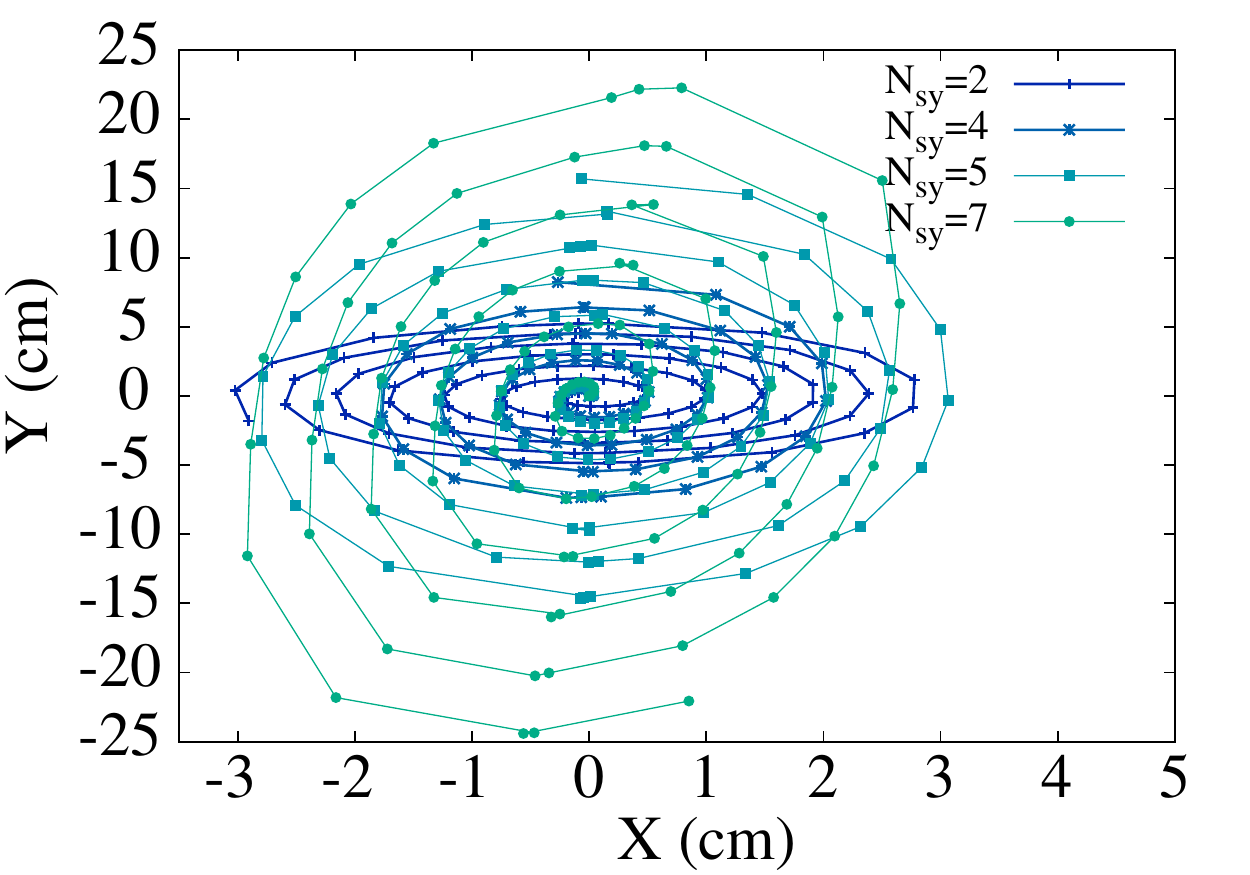}}
\caption{Plot of the jet profile projected on the $YZ$-plane (a) and $XY$-plane (b) for various relative intensities $N_{sy}$ of the noise along the $Y$-axis with respect to the noise $N_s$ on the $XY$-plane. $ V_0=\;10^4 $ V,  $ I_F=5 \;10^4 $, $N_s=10^{-3}$, $ h=20$ cm. The greater the asymmetry in the noise, the larger the asymmetry in the helix profile as visible in (b). Note the scales of X and Y axes are significantly different here. This transition is rather fast to the point that above an asymmetry ratio of $N_{sy}=7$ we could not get stable numerical solutions any more, and already at $N_{sy}=7$ the helix gets deformed towards 1D banana shape profiles (Fig. 5(a)).}
\label{FigNoiseFre-NY}
\end{figure}

We also investigated the effect of planar noise asymmetry and vertical driving force with a range of frequencies (see Figs.~\ref{FigExtension-N}-\ref{FigNoiseFre-NY}). A wide variety of beautiful helices is obtained in this way, confirming the richness of the phenomenology shown by the dynamics of electrified jets. We found that the planar driving force remains the best design principle. In fact, by introducing asymmetry in the planar noise we quickly move towards the one-dimensional scenario with sharper turns that locally thicken the fibre. We also found that an additional vertical driving force does not provide a significant reduction of the fibre thickness. In the future we plan to include an additional component of the driving force along the $Z$-axis and/or consider an oscillating external field in the model.

\subsection{Effect of the perturbation frequency}

\begin{figure}[ht]
\subfigure[]{\includegraphics[width=0.7\columnwidth]{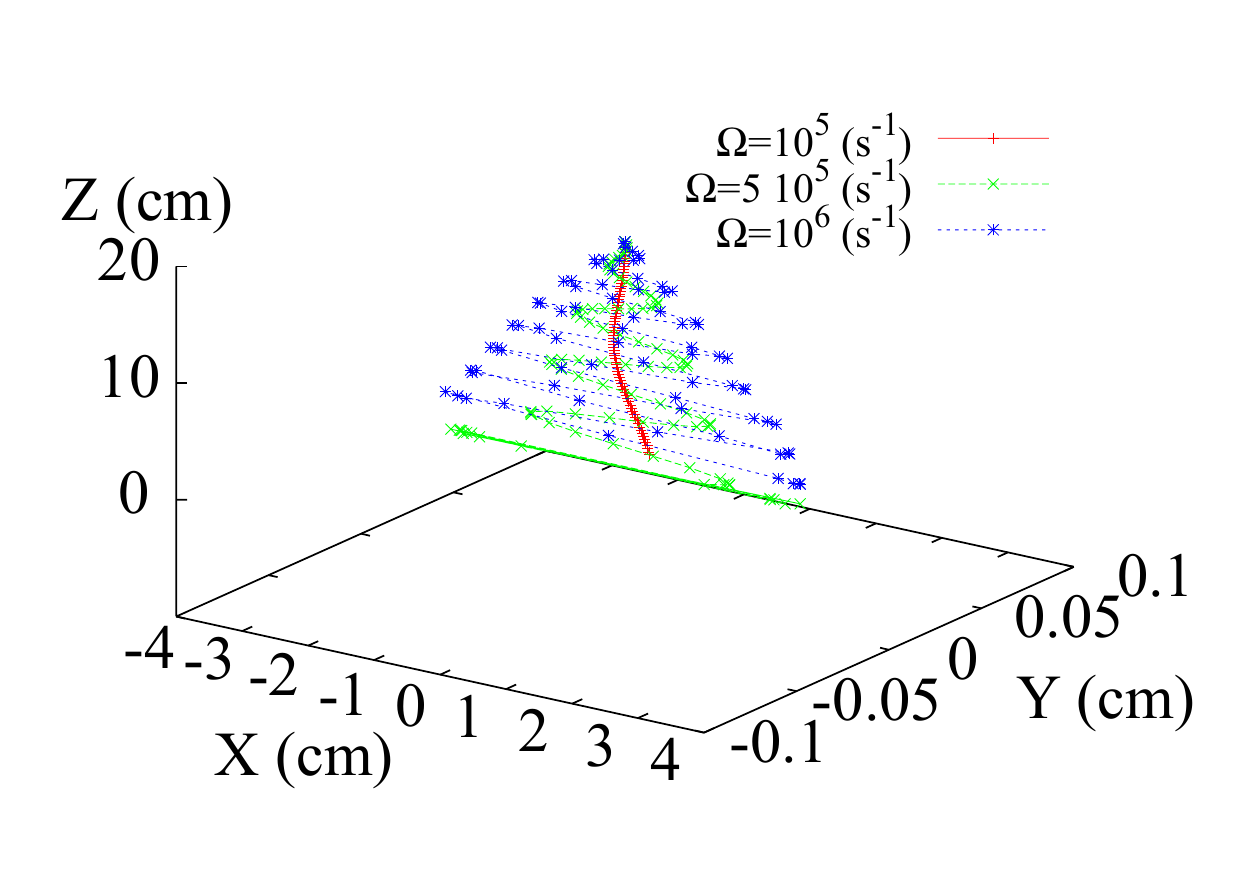}}
\subfigure[]{\includegraphics[width=0.7\columnwidth]{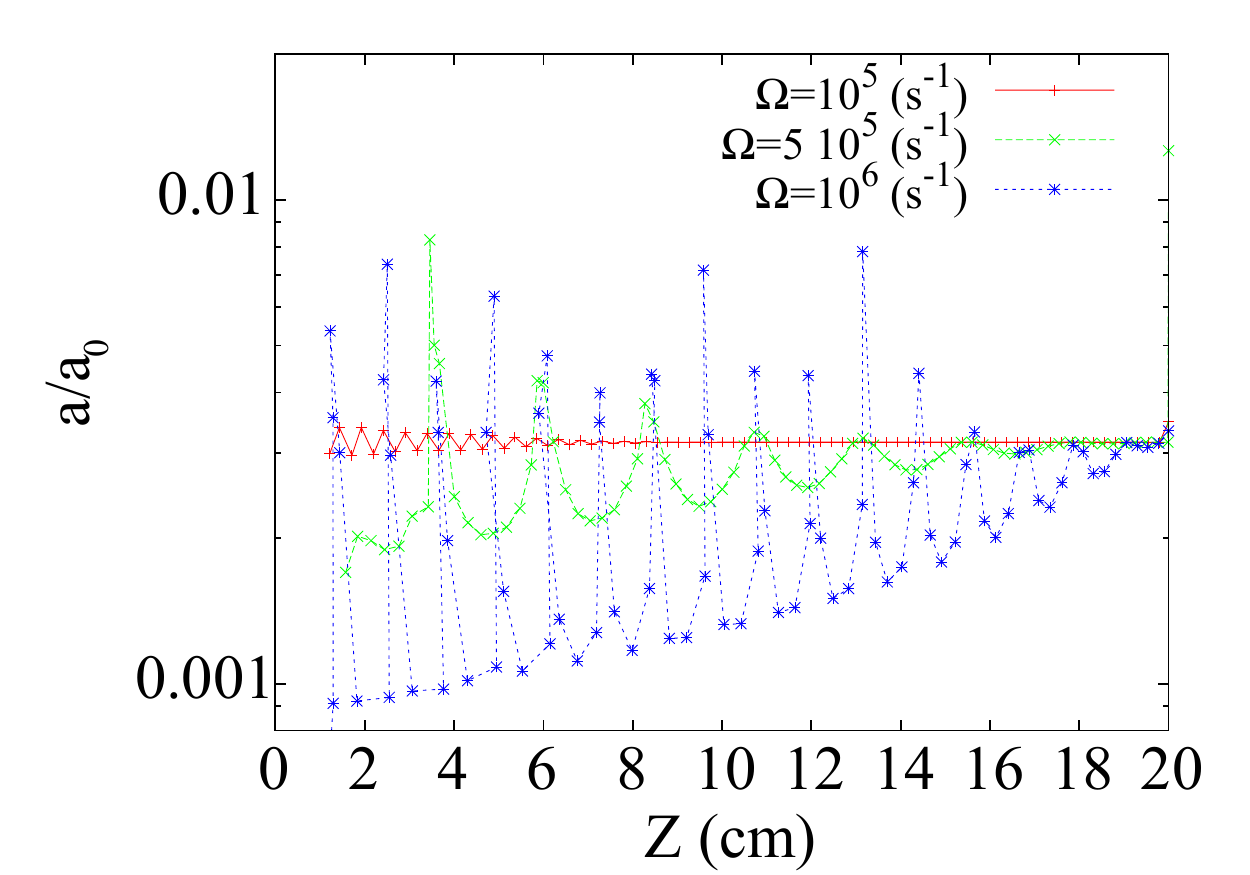}}
\caption{(a) The spiral structures for different values of the load frequency in the one-dimensional driving scenario along the $X$-axis. Parameters used for the simulations are $ V_0=10^4 $ V,  $ I_F=5 \;10^4 $, $N_s=10^{-3}$, $ h=20$ cm. (b) Corresponding fibre extension plot, where the characteristic spikes due to the local bending are visible. 
As for the circular case, also in this case the increase in frequency results in thinner fibres.}
\label{FigNoiseFre-1D}
\end{figure}

\begin{figure*}[ht]
\subfigure[]{\includegraphics[width=0.7\textwidth]{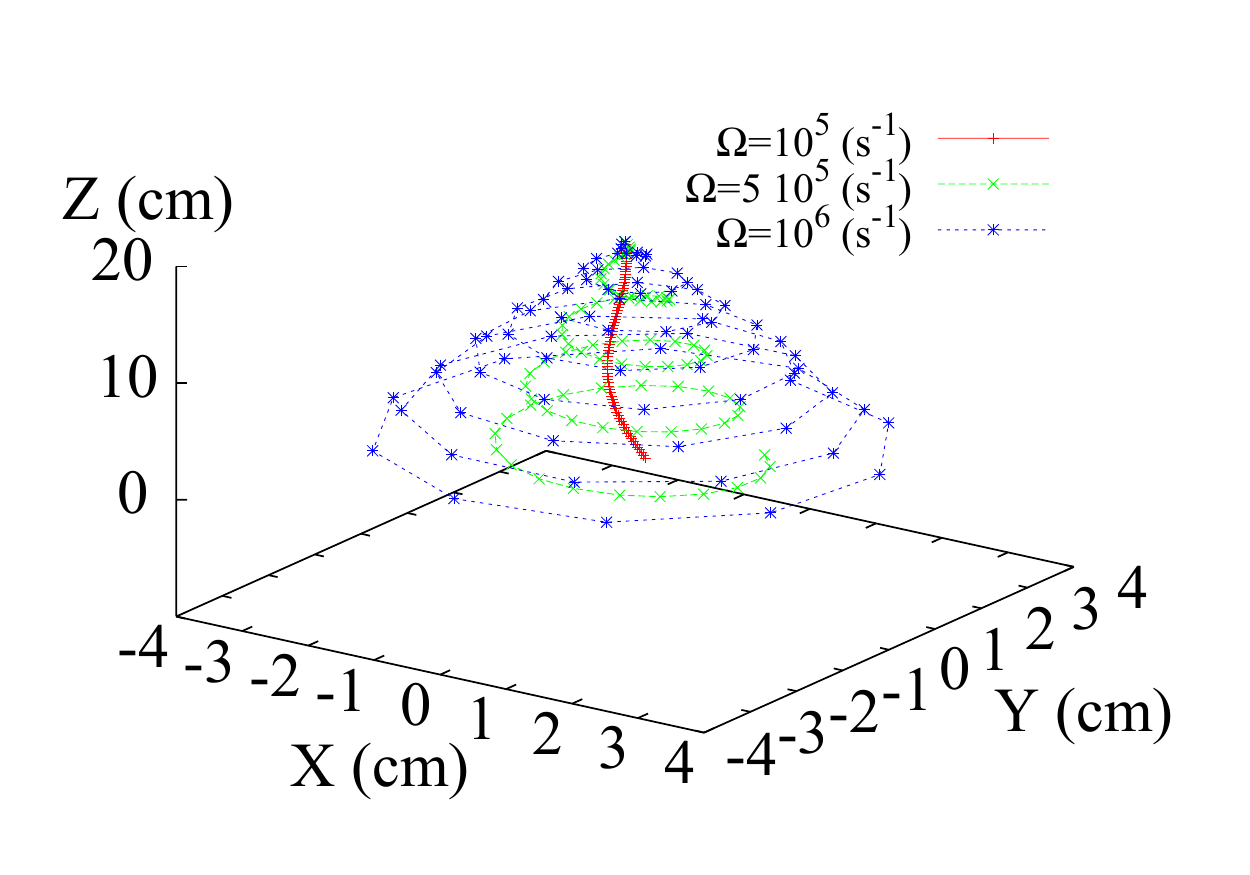}}
\subfigure[]{\includegraphics[width=0.7\textwidth]{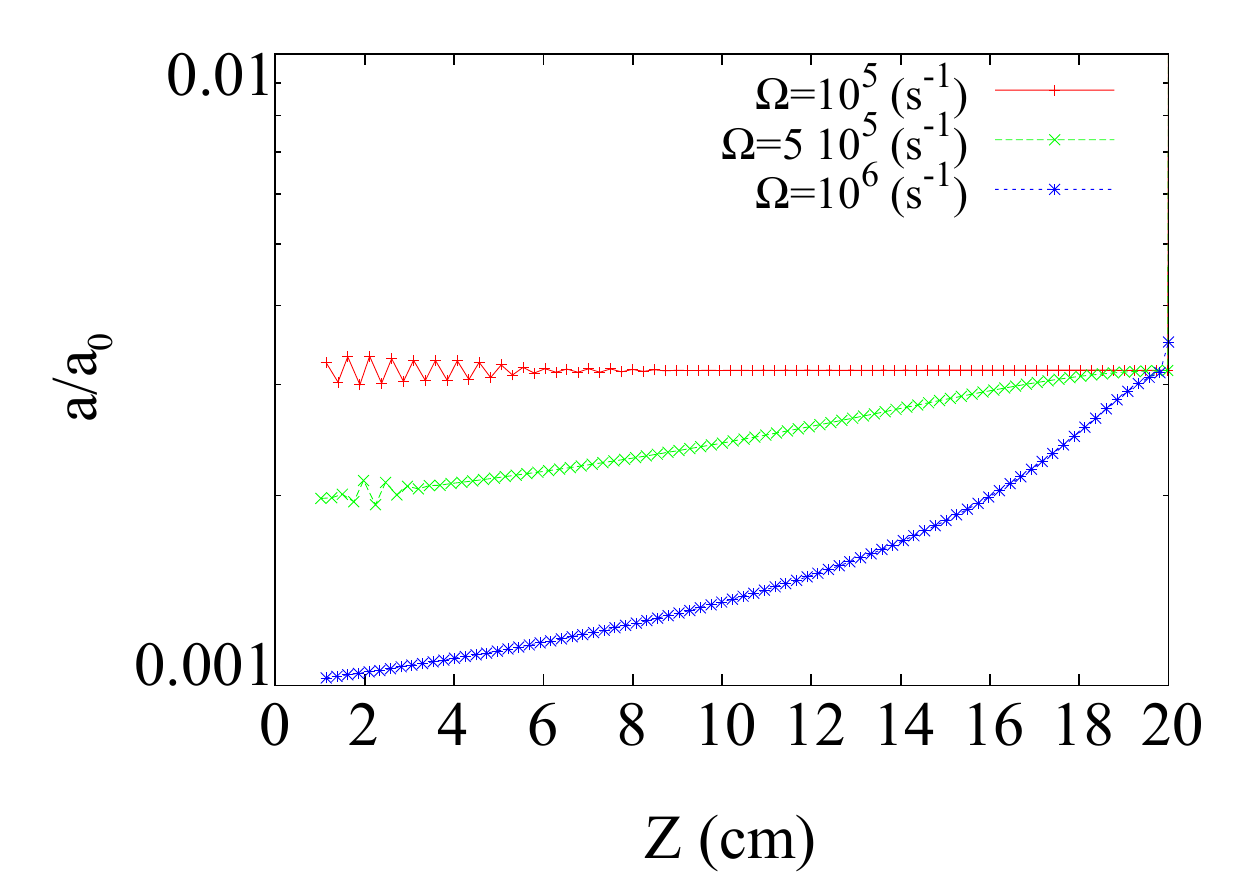}}
\caption{Comparison of the profiles obtained with different  values of the perturbation frequency $\Omega$. Parameters used for the simulations are as above except for $\Omega$. (a) Final configuration of the filament under a circular perturbation. (b) Plot of the elongation as a function of the distance from the collector. Similarly to what observed by varying the perturbation strength (Fig.~\ref{Figure-2})
also upon increasing the frequency $\Omega$ the radius of the spirals increases and stretches the fibre.}
\label{FigNoiseFre}
\end{figure*}

We have also explored the effect of the perturbation frequency, $\Omega$, by running
a series of simulations with $\Omega=10^4, 10^5, 5 \;10^5, 10^6\; \text{s}^{-1}$, for both planar and three-dimensional scenarios. The corresponding spiral structures are shown in Fig.~\ref{FigNoiseFre-1D} and in Fig.~\ref{FigNoiseFre}, respectively.
From these figures, it is apparent that the main effect of increasing the perturbation frequency is to produce
more compact and broader structures, i.e. spirals with a larger aperture and a shorter pitch.
This can be traced back to the fact that, by increasing the perturbation frequency, the 
Coulomb repulsion in the $XY$-plane is enhanced, thus leading to larger ratio between
the horizontal and vertical speeds, and hence a shorter pitch. Thus, increasing the frequency leads again to thinner fibres at 
the collector, offering an additional design parameter, which minimizes the fibre thickness.
To the best of our knowledge, such a design parameter has not been considered before.

\subsection{Effect of the location of the perturbation source}

In most experiments the polymer jet is seen to fall down along a straight-line configuration, prior to the development of bending instabilities. In other words, a vertical neck precedes the formation of spiral structures. Here, we have not been able to find any parameter regime for which the neck would smoothly develop into a spiral structure: the two configurations do not appear to belong together. What we have found instead is that spirals start precisely at the location  where the perturbation is applied, which means that when the perturbation is applied at the spinneret, no neck is observed.

\begin{figure*}[ht]
\subfigure[\label{Conf-K5}]{\includegraphics[width=0.7\textwidth]{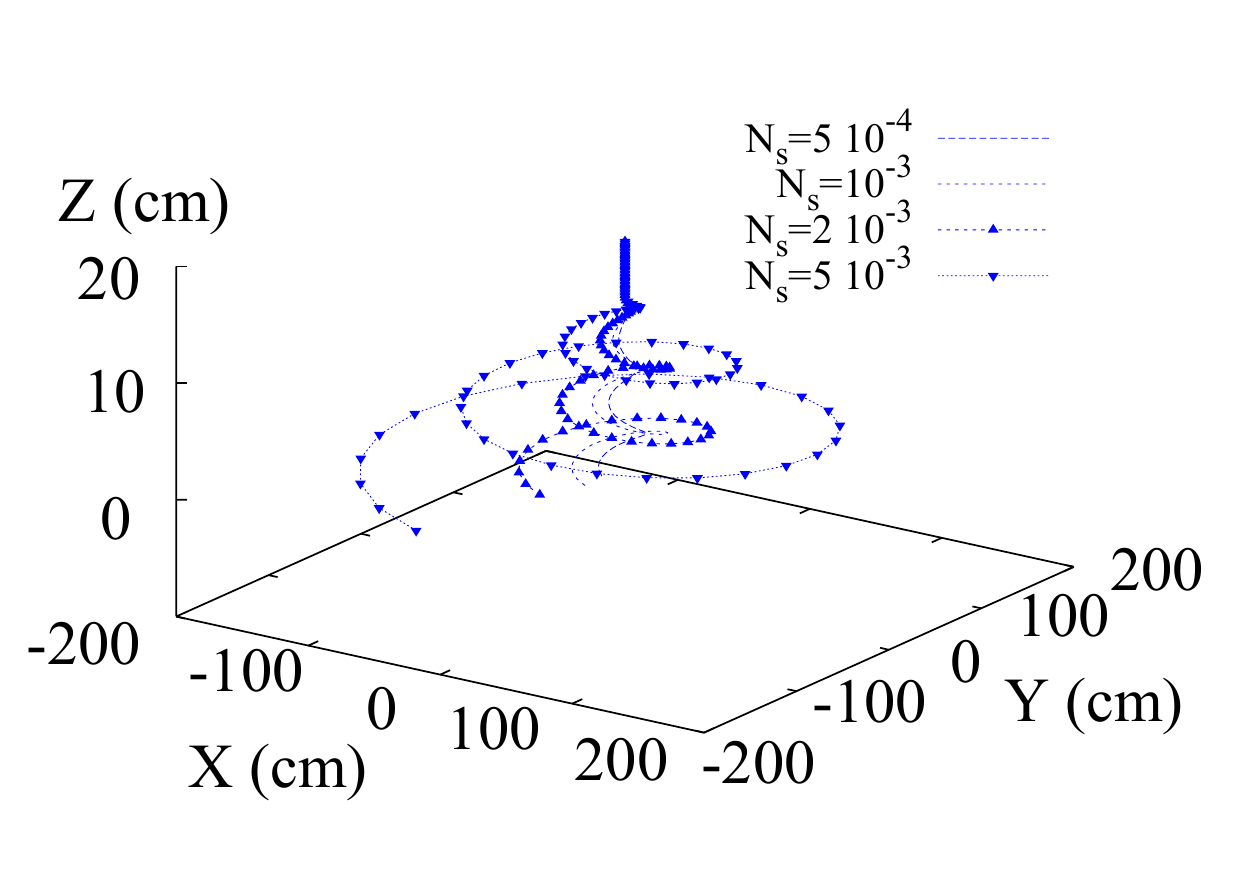}}
\subfigure[\label{Extension-K5}]{\includegraphics[width=0.7\textwidth]{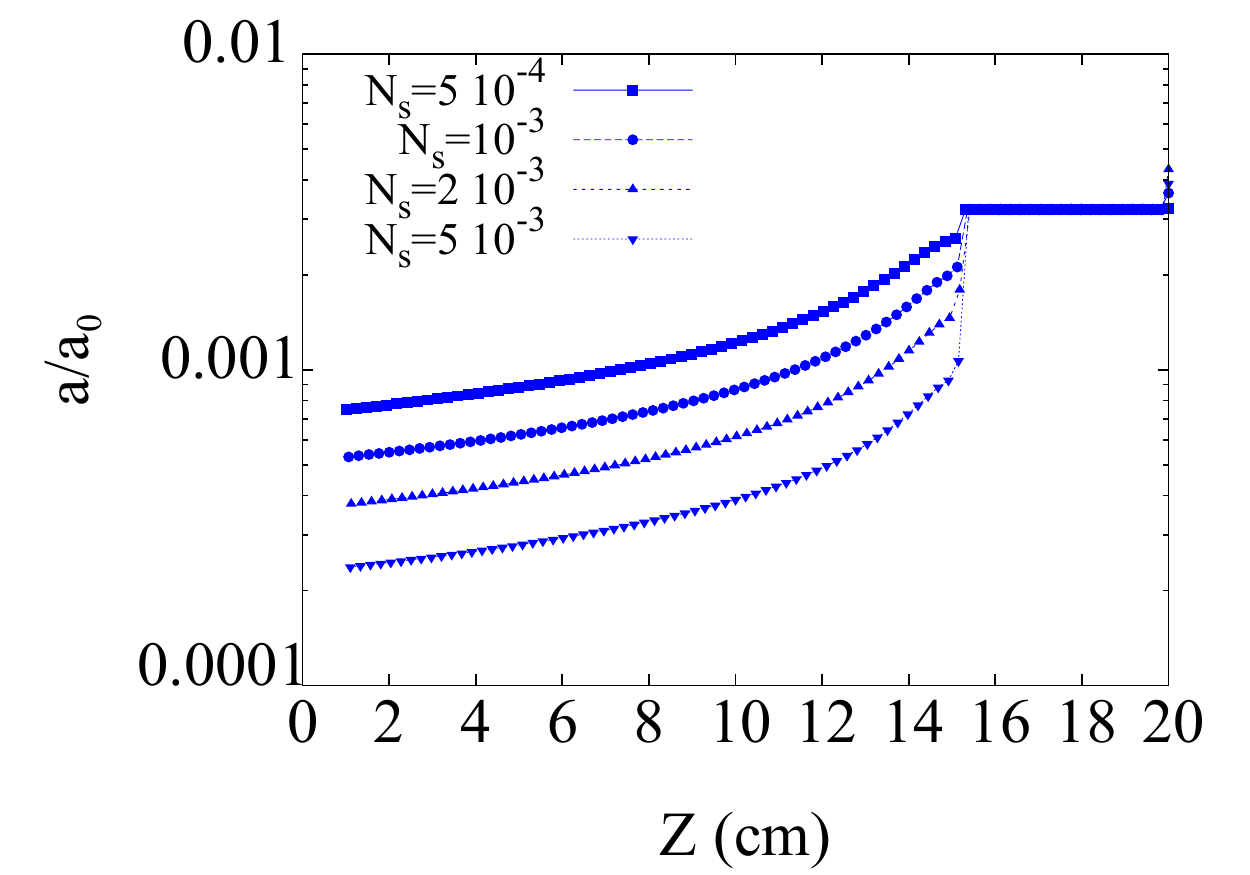}}
\caption{Effect on the spiral shape of the mechanical load applied $5$ cm below the pendant drop for various values of the load strength $N_s$. (a) Final configurations of the filament. Parameters are as above except for $N_s$. (b) Elongation as a function of the distance from the collector. Above the height of 15 cm, the location of the perturbation source, the fibre does not stretch appreciably. 
}\label{Figure-Neck-1}
\end{figure*}

To illustrate this point, in Fig.~\ref{Figure-Neck-1} we report the configuration
of the spiral and the thickness of the jet, for a case where the perturbation is applied 
$5$ cm below the pendant drop. The jet shows a clear neck, up precisely to the point where
the planar perturbation is applied.  
Subsequently, the spiral forms in the very same way as it did when 
the perturbation was placed at the pendant drop.
A series of simulations was carried out by changing the location of the perturbation, always to
find the same result: the spiral starts to develop precisely at the perturbation location. Furthermore, we have observed that, upon impinging on the collector, the
jet forms planar coil patterns, similar to those observed in experiments, 
as one would expect for a thread falling on a stationary surface 
\cite{Chiu-Webster2006}.

These findings suggest that the presence of the neck might reflect a different and more elaborate pathway to instability than just mechanical perturbation at the spinneret. Among others, a possibility is provided by environmental fluctuations, say electrostatic or perhaps hydrodynamic ones, which may require a finite waiting time before building up sufficient strength to trigger bending instabilities. 
The importance of the formation of the neck on the development of the instability has been discussed in the work of Li et al.~\cite{Li2013,Li2006} where a detailed analysis of the initial step of the spinning process was developed and linked to the physico-chemical properties of the polymer solution jet. The results presented in the work of Li et al. suggest a strong dependence of the instability on such parameters. In the present study, we have also considered a broad spectrum of different physical parameters with respect to the one that define the nondimensional quantities in Table~\ref{tab:Real_Values} (results not shown). For some values of the parameters we could not produce helices, suggesting that the properties of the polymer solutions strongly affect the instability process, hence in qualitative agreement with the results of Li et al.~\cite{Li2013,Li2006}. However, even when we did detect helices, we still could not observe any qualitative change of the picture presented in this work, namely that planar oscillations driving the spinneret offer a viable route to produce much thinner fibres, regardless of the chemical physical properties of the polymer solution. A detailed analysis of the effect of the solutions used in the experiments will make the object of future work.


\section{Conclusions}

Summarizing, we have presented a computational study of the effects of a fast-oscillating perturbation on the formation and dynamics of spiral structures in electrified jets used for electrospinning experiments. In particular, we have provided numerical evidence of a direct dependence of the spiral aperture on the perturbation amplitude and frequency. As a result, our simulations suggest that both parameters could be tuned in order to minimize the fibre thickness at the collector, by maximizing the spiral length. This might open up optimal design protocols in electrospinning experiments. In other words, independently of the  rheology of the polymer solution, provided that fibres can be electrospun, applying a driving perturbation oscillating at high frequency at the pendant drop can further reduce the thickness.


\subsection{Acknowledgements}
M. Bernaschi and U. Amato are kindly acknowledged for computational help with the project.
The research leading to these results has received funding from the European Research Council under the European Union's Seventh Framework Programme (FP/2007-2013)/ERC Grant Agreement n. 306357 (ERC Starting Grant ``NANO-JETS''). 
One of the authors (SS) wishes to acknowledge the Erwin Schroedinger Institute (ESI), for kind hospitality
and financial support under the ESI Senior Fellow program. 

\bibliography{library}

\begin{thebibliography}{42}%
\makeatletter
\providecommand \@ifxundefined [1]{%
 \@ifx{#1\undefined}
}%
\providecommand \@ifnum [1]{%
 \ifnum #1\expandafter \@firstoftwo
 \else \expandafter \@secondoftwo
 \fi
}%
\providecommand \@ifx [1]{%
 \ifx #1\expandafter \@firstoftwo
 \else \expandafter \@secondoftwo
 \fi
}%
\providecommand \natexlab [1]{#1}%
\providecommand \enquote  [1]{``#1''}%
\providecommand \bibnamefont  [1]{#1}%
\providecommand \bibfnamefont [1]{#1}%
\providecommand \citenamefont [1]{#1}%
\providecommand \href@noop [0]{\@secondoftwo}%
\providecommand \href [0]{\begingroup \@sanitize@url \@href}%
\providecommand \@href[1]{\@@startlink{#1}\@@href}%
\providecommand \@@href[1]{\endgroup#1\@@endlink}%
\providecommand \@sanitize@url [0]{\catcode `\\12\catcode `\$12\catcode
  `\&12\catcode `\#12\catcode `\^12\catcode `\_12\catcode `\%12\relax}%
\providecommand \@@startlink[1]{}%
\providecommand \@@endlink[0]{}%
\providecommand \url  [0]{\begingroup\@sanitize@url \@url }%
\providecommand \@url [1]{\endgroup\@href {#1}{\urlprefix }}%
\providecommand \urlprefix  [0]{URL }%
\providecommand \Eprint [0]{\href }%
\providecommand \doibase [0]{http://dx.doi.org/}%
\providecommand \selectlanguage [0]{\@gobble}%
\providecommand \bibinfo  [0]{\@secondoftwo}%
\providecommand \bibfield  [0]{\@secondoftwo}%
\providecommand \translation [1]{[#1]}%
\providecommand \BibitemOpen [0]{}%
\providecommand \bibitemStop [0]{}%
\providecommand \bibitemNoStop [0]{.\EOS\space}%
\providecommand \EOS [0]{\spacefactor3000\relax}%
\providecommand \BibitemShut  [1]{\csname bibitem#1\endcsname}%
\let\auto@bib@innerbib\@empty
\bibitem [{\citenamefont {Yarin}(1993)}]{Yarin}%
  \BibitemOpen
  \bibfield  {author} {\bibinfo {author} {\bibfnamefont {Alexander~L}\
  \bibnamefont {Yarin}},\ }\href@noop {} {\emph {\bibinfo {title} {{Free liquid
  jets and films: Hydrodynamics and rheology}}}}\ (\bibinfo  {publisher}
  {Longman Scientific \& Technical (New York)},\ \bibinfo {year}
  {1993})\BibitemShut {NoStop}%
\bibitem [{\citenamefont {Reneker}\ \emph {et~al.}(2000)\citenamefont
  {Reneker}, \citenamefont {Yarin}, \citenamefont {Fong},\ and\ \citenamefont
  {Koombhongse}}]{Reneker2000}%
  \BibitemOpen
  \bibfield  {author} {\bibinfo {author} {\bibfnamefont {Darrell~H.}\
  \bibnamefont {Reneker}}, \bibinfo {author} {\bibfnamefont {Alexander~L.}\
  \bibnamefont {Yarin}}, \bibinfo {author} {\bibfnamefont {Hao}\ \bibnamefont
  {Fong}}, \ and\ \bibinfo {author} {\bibfnamefont {Sureeporn}\ \bibnamefont
  {Koombhongse}},\ }\bibfield  {title} {\enquote {\bibinfo {title} {{Bending
  instability of electrically charged liquid jets of polymer solutions in
  electrospinning}},}\ }\href {\doibase 10.1063/1.373532} {\bibfield  {journal}
  {\bibinfo  {journal} {Journal of Applied Physics}\ }\textbf {\bibinfo
  {volume} {87}},\ \bibinfo {pages} {4531} (\bibinfo {year}
  {2000})}\BibitemShut {NoStop}%
\bibitem [{\citenamefont {Yarin}\ \emph {et~al.}(2001)\citenamefont {Yarin},
  \citenamefont {Koombhongse},\ and\ \citenamefont {Reneker}}]{Yarin2001}%
  \BibitemOpen
  \bibfield  {author} {\bibinfo {author} {\bibfnamefont {A.~L.}\ \bibnamefont
  {Yarin}}, \bibinfo {author} {\bibfnamefont {S.}~\bibnamefont {Koombhongse}},
  \ and\ \bibinfo {author} {\bibfnamefont {D.~H.}\ \bibnamefont {Reneker}},\
  }\bibfield  {title} {\enquote {\bibinfo {title} {{Taylor cone and jetting
  from liquid droplets in electrospinning of nanofibers}},}\ }\href {\doibase
  10.1063/1.1408260} {\bibfield  {journal} {\bibinfo  {journal} {Journal of
  Applied Physics}\ }\textbf {\bibinfo {volume} {90}},\ \bibinfo {pages} {4836}
  (\bibinfo {year} {2001})}\BibitemShut {NoStop}%
\bibitem [{\citenamefont {Loscertales}\ \emph {et~al.}(2002)\citenamefont
  {Loscertales}, \citenamefont {Barrero}, \citenamefont {Guerrero},
  \citenamefont {Cortijo}, \citenamefont {Marquez},\ and\ \citenamefont
  {Ga\~{n}\'{a}n Calvo}}]{Loscertales2002}%
  \BibitemOpen
  \bibfield  {author} {\bibinfo {author} {\bibfnamefont {I~G}\ \bibnamefont
  {Loscertales}}, \bibinfo {author} {\bibfnamefont {A}~\bibnamefont {Barrero}},
  \bibinfo {author} {\bibfnamefont {I}~\bibnamefont {Guerrero}}, \bibinfo
  {author} {\bibfnamefont {R}~\bibnamefont {Cortijo}}, \bibinfo {author}
  {\bibfnamefont {M}~\bibnamefont {Marquez}}, \ and\ \bibinfo {author}
  {\bibfnamefont {A~M}\ \bibnamefont {Ga\~{n}\'{a}n Calvo}},\ }\bibfield
  {title} {\enquote {\bibinfo {title} {{Micro/nano encapsulation via
  electrified coaxial liquid jets.}}}\ }\href {\doibase
  10.1126/science.1067595} {\bibfield  {journal} {\bibinfo  {journal} {Science
  (New York, N.Y.)}\ }\textbf {\bibinfo {volume} {295}},\ \bibinfo {pages}
  {1695--8} (\bibinfo {year} {2002})}\BibitemShut {NoStop}%
\bibitem [{\citenamefont {Dzenis}(2004)}]{Dzenis2004}%
  \BibitemOpen
  \bibfield  {author} {\bibinfo {author} {\bibfnamefont {Yuris~A}\ \bibnamefont
  {Dzenis}},\ }\bibfield  {title} {\enquote {\bibinfo {title} {{Spinning
  Continuous Fibers for Nanotechnology Spinning Continuous Fibers for
  Nanotechnology}},}\ }\href@noop {} {\bibfield  {journal} {\bibinfo  {journal}
  {Science}\ }\textbf {\bibinfo {volume} {304}},\ \bibinfo {pages} {1917--1919}
  (\bibinfo {year} {2004})}\BibitemShut {NoStop}%
\bibitem [{\citenamefont {Helgeson}\ \emph {et~al.}(2008)\citenamefont
  {Helgeson}, \citenamefont {Grammatikos}, \citenamefont {Deitzel},\ and\
  \citenamefont {Wagner}}]{Helgeson2008}%
  \BibitemOpen
  \bibfield  {author} {\bibinfo {author} {\bibfnamefont {Matthew~E.}\
  \bibnamefont {Helgeson}}, \bibinfo {author} {\bibfnamefont {Kristie~N.}\
  \bibnamefont {Grammatikos}}, \bibinfo {author} {\bibfnamefont {Joseph~M.}\
  \bibnamefont {Deitzel}}, \ and\ \bibinfo {author} {\bibfnamefont {Norman~J.}\
  \bibnamefont {Wagner}},\ }\bibfield  {title} {\enquote {\bibinfo {title}
  {{Theory and kinematic measurements of the mechanics of stable electrospun
  polymer jets}},}\ }\href {\doibase 10.1016/j.polymer.2008.04.025} {\bibfield
  {journal} {\bibinfo  {journal} {Polymer}\ }\textbf {\bibinfo {volume} {49}},\
  \bibinfo {pages} {2924--2936} (\bibinfo {year} {2008})}\BibitemShut {NoStop}%
\bibitem [{\citenamefont {Thompson}\ \emph {et~al.}(2007)\citenamefont
  {Thompson}, \citenamefont {Chase}, \citenamefont {Yarin},\ and\ \citenamefont
  {Reneker}}]{Thompson2007}%
  \BibitemOpen
  \bibfield  {author} {\bibinfo {author} {\bibfnamefont {C.~J.}\ \bibnamefont
  {Thompson}}, \bibinfo {author} {\bibfnamefont {G.~G.}\ \bibnamefont {Chase}},
  \bibinfo {author} {\bibfnamefont {A.~L.}\ \bibnamefont {Yarin}}, \ and\
  \bibinfo {author} {\bibfnamefont {D.H.}\ \bibnamefont {Reneker}},\ }\bibfield
   {title} {\enquote {\bibinfo {title} {{Effects of parameters on nanofiber
  diameter determined from electrospinning model}},}\ }\href {\doibase
  10.1016/j.polymer.2007.09.017} {\bibfield  {journal} {\bibinfo  {journal}
  {Polymer}\ }\textbf {\bibinfo {volume} {48}},\ \bibinfo {pages} {6913--6922}
  (\bibinfo {year} {2007})}\BibitemShut {NoStop}%
\bibitem [{\citenamefont {Grafahrend}\ \emph {et~al.}(2011)\citenamefont
  {Grafahrend}, \citenamefont {Heffels}, \citenamefont {Beer}, \citenamefont
  {Gasteier}, \citenamefont {M\"{o}ller}, \citenamefont {Boehm}, \citenamefont
  {Dalton},\ and\ \citenamefont {Groll}}]{Grafahrend2011}%
  \BibitemOpen
  \bibfield  {author} {\bibinfo {author} {\bibfnamefont {Dirk}\ \bibnamefont
  {Grafahrend}}, \bibinfo {author} {\bibfnamefont {Karl-Heinz}\ \bibnamefont
  {Heffels}}, \bibinfo {author} {\bibfnamefont {Meike~V}\ \bibnamefont {Beer}},
  \bibinfo {author} {\bibfnamefont {Peter}\ \bibnamefont {Gasteier}}, \bibinfo
  {author} {\bibfnamefont {Martin}\ \bibnamefont {M\"{o}ller}}, \bibinfo
  {author} {\bibfnamefont {Gabriele}\ \bibnamefont {Boehm}}, \bibinfo {author}
  {\bibfnamefont {Paul~D}\ \bibnamefont {Dalton}}, \ and\ \bibinfo {author}
  {\bibfnamefont {J\"{u}rgen}\ \bibnamefont {Groll}},\ }\bibfield  {title}
  {\enquote {\bibinfo {title} {{Degradable polyester scaffolds with controlled
  surface chemistry combining minimal protein adsorption with specific
  bioactivation.}}}\ }\href {\doibase 10.1038/nmat2904} {\bibfield  {journal}
  {\bibinfo  {journal} {Nature materials}\ }\textbf {\bibinfo {volume} {10}},\
  \bibinfo {pages} {67--73} (\bibinfo {year} {2011})}\BibitemShut {NoStop}%
\bibitem [{\citenamefont {Greenfeld}\ \emph {et~al.}(2012)\citenamefont
  {Greenfeld}, \citenamefont {Fezzaa}, \citenamefont {Rafailovich},\ and\
  \citenamefont {Zussman}}]{Greenfeld2012}%
  \BibitemOpen
  \bibfield  {author} {\bibinfo {author} {\bibfnamefont {Israel}\ \bibnamefont
  {Greenfeld}}, \bibinfo {author} {\bibfnamefont {Kamel}\ \bibnamefont
  {Fezzaa}}, \bibinfo {author} {\bibfnamefont {Miriam~H.}\ \bibnamefont
  {Rafailovich}}, \ and\ \bibinfo {author} {\bibfnamefont {Eyal}\ \bibnamefont
  {Zussman}},\ }\bibfield  {title} {\enquote {\bibinfo {title} {{Fast X-ray
  Phase-Contrast Imaging of Electrospinning Polymer Jets: Measurements of
  Radius, Velocity, and Concentration}},}\ }\href {\doibase 10.1021/ma300237j}
  {\bibfield  {journal} {\bibinfo  {journal} {Macromolecules}\ }\textbf
  {\bibinfo {volume} {45}},\ \bibinfo {pages} {3616--3626} (\bibinfo {year}
  {2012})}\BibitemShut {NoStop}%
\bibitem [{\citenamefont {Onses}\ \emph {et~al.}(2013)\citenamefont {Onses},
  \citenamefont {Song}, \citenamefont {Williamson}, \citenamefont {Sutanto},
  \citenamefont {Ferreira}, \citenamefont {Alleyne}, \citenamefont {Nealey},
  \citenamefont {Ahn},\ and\ \citenamefont {Rogers}}]{Onses2013}%
  \BibitemOpen
  \bibfield  {author} {\bibinfo {author} {\bibfnamefont {M.~Serdar}\
  \bibnamefont {Onses}}, \bibinfo {author} {\bibfnamefont {Chiho}\ \bibnamefont
  {Song}}, \bibinfo {author} {\bibfnamefont {Lance}\ \bibnamefont
  {Williamson}}, \bibinfo {author} {\bibfnamefont {Erick}\ \bibnamefont
  {Sutanto}}, \bibinfo {author} {\bibfnamefont {Placid~M.}\ \bibnamefont
  {Ferreira}}, \bibinfo {author} {\bibfnamefont {Andrew~G.}\ \bibnamefont
  {Alleyne}}, \bibinfo {author} {\bibfnamefont {Paul~F.}\ \bibnamefont
  {Nealey}}, \bibinfo {author} {\bibfnamefont {Heejoon}\ \bibnamefont {Ahn}}, \
  and\ \bibinfo {author} {\bibfnamefont {John~A.}\ \bibnamefont {Rogers}},\
  }\bibfield  {title} {\enquote {\bibinfo {title} {{Hierarchical patterns of
  three-dimensional block-copolymer films formed by electrohydrodynamic jet
  printing and self-assembly.}}}\ }\href {\doibase 10.1038/nnano.2013.160}
  {\bibfield  {journal} {\bibinfo  {journal} {Nature nanotechnology}\ }\textbf
  {\bibinfo {volume} {8}},\ \bibinfo {pages} {667--75} (\bibinfo {year}
  {2013})}\BibitemShut {NoStop}%
\bibitem [{\citenamefont {Min}\ \emph {et~al.}(2013)\citenamefont {Min},
  \citenamefont {Kim}, \citenamefont {Kim}, \citenamefont {Cho}, \citenamefont
  {Noh}, \citenamefont {Yang}, \citenamefont {Cho},\ and\ \citenamefont
  {Lee}}]{Min2013}%
  \BibitemOpen
  \bibfield  {author} {\bibinfo {author} {\bibfnamefont {Sung-Yong}\
  \bibnamefont {Min}}, \bibinfo {author} {\bibfnamefont {Tae-Sik}\ \bibnamefont
  {Kim}}, \bibinfo {author} {\bibfnamefont {Beom~Joon}\ \bibnamefont {Kim}},
  \bibinfo {author} {\bibfnamefont {Himchan}\ \bibnamefont {Cho}}, \bibinfo
  {author} {\bibfnamefont {Yong-Young}\ \bibnamefont {Noh}}, \bibinfo {author}
  {\bibfnamefont {Hoichang}\ \bibnamefont {Yang}}, \bibinfo {author}
  {\bibfnamefont {Jeong~Ho}\ \bibnamefont {Cho}}, \ and\ \bibinfo {author}
  {\bibfnamefont {Tae-Woo}\ \bibnamefont {Lee}},\ }\bibfield  {title} {\enquote
  {\bibinfo {title} {{Large-scale organic nanowire lithography and
  electronics.}}}\ }\href {\doibase 10.1038/ncomms2785} {\bibfield  {journal}
  {\bibinfo  {journal} {Nature communications}\ }\textbf {\bibinfo {volume}
  {4}},\ \bibinfo {pages} {1773} (\bibinfo {year} {2013})}\BibitemShut
  {NoStop}%
\bibitem [{\citenamefont {Regev}\ \emph {et~al.}(2013)\citenamefont {Regev},
  \citenamefont {Arinstein},\ and\ \citenamefont {Zussman}}]{Regev2013}%
  \BibitemOpen
  \bibfield  {author} {\bibinfo {author} {\bibfnamefont {Omri}\ \bibnamefont
  {Regev}}, \bibinfo {author} {\bibfnamefont {Arkadii}\ \bibnamefont
  {Arinstein}}, \ and\ \bibinfo {author} {\bibfnamefont {Eyal}\ \bibnamefont
  {Zussman}},\ }\bibfield  {title} {\enquote {\bibinfo {title} {{Creep anomaly
  in electrospun fibers made of globular proteins}},}\ }\href {\doibase
  10.1103/PhysRevE.88.062605} {\bibfield  {journal} {\bibinfo  {journal}
  {Physical Review E}\ }\textbf {\bibinfo {volume} {88}},\ \bibinfo {pages}
  {062605} (\bibinfo {year} {2013})}\BibitemShut {NoStop}%
\bibitem [{\citenamefont {Nakano}\ \emph {et~al.}(2012)\citenamefont {Nakano},
  \citenamefont {Miki}, \citenamefont {Hishida},\ and\ \citenamefont
  {Hotta}}]{Nakano2012}%
  \BibitemOpen
  \bibfield  {author} {\bibinfo {author} {\bibfnamefont {Atsushi}\ \bibnamefont
  {Nakano}}, \bibinfo {author} {\bibfnamefont {Norihisa}\ \bibnamefont {Miki}},
  \bibinfo {author} {\bibfnamefont {Koichi}\ \bibnamefont {Hishida}}, \ and\
  \bibinfo {author} {\bibfnamefont {Atsushi}\ \bibnamefont {Hotta}},\
  }\bibfield  {title} {\enquote {\bibinfo {title} {{Solution parameters for the
  fabrication of thinner silicone fibers by electrospinning}},}\ }\href
  {\doibase 10.1103/PhysRevE.86.011801} {\bibfield  {journal} {\bibinfo
  {journal} {Physical Review E}\ }\textbf {\bibinfo {volume} {86}},\ \bibinfo
  {pages} {011801} (\bibinfo {year} {2012})}\BibitemShut {NoStop}%
\bibitem [{\citenamefont {Greiner}\ and\ \citenamefont
  {Wendorff}(2007)}]{Greiner2007}%
  \BibitemOpen
  \bibfield  {author} {\bibinfo {author} {\bibfnamefont {Andreas}\ \bibnamefont
  {Greiner}}\ and\ \bibinfo {author} {\bibfnamefont {Joachim~H}\ \bibnamefont
  {Wendorff}},\ }\bibfield  {title} {\enquote {\bibinfo {title}
  {{Electrospinning: a fascinating method for the preparation of ultrathin
  fibers.}}}\ }\href {\doibase 10.1002/anie.200604646} {\bibfield  {journal}
  {\bibinfo  {journal} {Angewandte Chemie (International ed. in English)}\
  }\textbf {\bibinfo {volume} {46}},\ \bibinfo {pages} {5670--703} (\bibinfo
  {year} {2007})}\BibitemShut {NoStop}%
\bibitem [{\citenamefont {Fridrikh}\ \emph {et~al.}(2003)\citenamefont
  {Fridrikh}, \citenamefont {Yu}, \citenamefont {Brenner},\ and\ \citenamefont
  {Rutledge}}]{Fridrikh2003}%
  \BibitemOpen
  \bibfield  {author} {\bibinfo {author} {\bibfnamefont {Sergey}\ \bibnamefont
  {Fridrikh}}, \bibinfo {author} {\bibfnamefont {Jian}\ \bibnamefont {Yu}},
  \bibinfo {author} {\bibfnamefont {Michael}\ \bibnamefont {Brenner}}, \ and\
  \bibinfo {author} {\bibfnamefont {Gregory}\ \bibnamefont {Rutledge}},\
  }\bibfield  {title} {\enquote {\bibinfo {title} {{Controlling the Fiber
  Diameter during Electrospinning}},}\ }\href {\doibase
  10.1103/PhysRevLett.90.144502} {\bibfield  {journal} {\bibinfo  {journal}
  {Physical Review Letters}\ }\textbf {\bibinfo {volume} {90}},\ \bibinfo
  {pages} {144502} (\bibinfo {year} {2003})}\BibitemShut {NoStop}%
\bibitem [{\citenamefont {Camposeo}\ \emph {et~al.}(2013)\citenamefont
  {Camposeo}, \citenamefont {Persano},\ and\ \citenamefont
  {Pisignano}}]{Camposeo2013}%
  \BibitemOpen
  \bibfield  {author} {\bibinfo {author} {\bibfnamefont {Andrea}\ \bibnamefont
  {Camposeo}}, \bibinfo {author} {\bibfnamefont {Luana}\ \bibnamefont
  {Persano}}, \ and\ \bibinfo {author} {\bibfnamefont {Dario}\ \bibnamefont
  {Pisignano}},\ }\bibfield  {title} {\enquote {\bibinfo {title}
  {{Light-Emitting Electrospun Nanofibers for Nanophotonics and
  Optoelectronics}},}\ }\href {\doibase 10.1002/mame.201200277} {\bibfield
  {journal} {\bibinfo  {journal} {Macromolecular Materials and Engineering}\
  }\textbf {\bibinfo {volume} {298}},\ \bibinfo {pages} {487--503} (\bibinfo
  {year} {2013})}\BibitemShut {NoStop}%
\bibitem [{\citenamefont {Pisignano}(2013)}]{Pisignano2013}%
  \BibitemOpen
  \bibfield  {author} {\bibinfo {author} {\bibfnamefont {Dario}\ \bibnamefont
  {Pisignano}},\ }\href@noop {} {\emph {\bibinfo {title} {{Polymer Nanofibers:
  Building Blocks for Nanotechnology}}}},\ \bibinfo {number} {29}\ (\bibinfo
  {publisher} {Royal Society of Chemistry},\ \bibinfo {year}
  {2013})\BibitemShut {NoStop}%
\bibitem [{\citenamefont {Taylor}(1969)}]{Taylor1969}%
  \BibitemOpen
  \bibfield  {author} {\bibinfo {author} {\bibfnamefont {G.}~\bibnamefont
  {Taylor}},\ }\bibfield  {title} {\enquote {\bibinfo {title} {{Electrically
  Driven Jets}},}\ }\href {\doibase 10.1098/rspa.1969.0205} {\bibfield
  {journal} {\bibinfo  {journal} {Proceedings of the Royal Society A:
  Mathematical, Physical and Engineering Sciences}\ }\textbf {\bibinfo {volume}
  {313}},\ \bibinfo {pages} {453--475} (\bibinfo {year} {1969})}\BibitemShut
  {NoStop}%
\bibitem [{\citenamefont {Marin}\ \emph {et~al.}(2008)\citenamefont {Marin},
  \citenamefont {Riboux}, \citenamefont {Loscertales},\ and\ \citenamefont
  {Barrero}}]{Marin2008}%
  \BibitemOpen
  \bibfield  {author} {\bibinfo {author} {\bibfnamefont {Alvaro~G.}\
  \bibnamefont {Marin}}, \bibinfo {author} {\bibfnamefont {Guillaume}\
  \bibnamefont {Riboux}}, \bibinfo {author} {\bibfnamefont {Ignacio~G}\
  \bibnamefont {Loscertales}}, \ and\ \bibinfo {author} {\bibfnamefont
  {Antonio}\ \bibnamefont {Barrero}},\ }\bibfield  {title} {\enquote {\bibinfo
  {title} {{Whipping Instabilities in Electrified Liquid Jets}},}\ }\href@noop
  {} {\ ,\ \bibinfo {pages} {3} (\bibinfo {year} {2008})},\ \Eprint
  {http://arxiv.org/abs/0810.0155} {arXiv:0810.0155} \BibitemShut {NoStop}%
\bibitem [{\citenamefont {Liu}\ \emph {et~al.}(2005)\citenamefont {Liu},
  \citenamefont {Reccius},\ and\ \citenamefont {Craighead}}]{Liu2005}%
  \BibitemOpen
  \bibfield  {author} {\bibinfo {author} {\bibfnamefont {Haiqing}\ \bibnamefont
  {Liu}}, \bibinfo {author} {\bibfnamefont {Christian~H.}\ \bibnamefont
  {Reccius}}, \ and\ \bibinfo {author} {\bibfnamefont {H.~G.}\ \bibnamefont
  {Craighead}},\ }\bibfield  {title} {\enquote {\bibinfo {title} {{Single
  electrospun regioregular poly(3-hexylthiophene) nanofiber field-effect
  transistor}},}\ }\href {\doibase 10.1063/1.2149980} {\bibfield  {journal}
  {\bibinfo  {journal} {Applied Physics Letters}\ }\textbf {\bibinfo {volume}
  {87}},\ \bibinfo {pages} {253106} (\bibinfo {year} {2005})}\BibitemShut
  {NoStop}%
\bibitem [{\citenamefont {Persano}\ \emph {et~al.}(2014)\citenamefont
  {Persano}, \citenamefont {Camposeo}, \citenamefont {Carro}, \citenamefont
  {Fasano}, \citenamefont {Moffa}, \citenamefont {Manco}, \citenamefont
  {D'Agostino},\ and\ \citenamefont {Pisignano}}]{Persano2014}%
  \BibitemOpen
  \bibfield  {author} {\bibinfo {author} {\bibfnamefont {Luana}\ \bibnamefont
  {Persano}}, \bibinfo {author} {\bibfnamefont {Andrea}\ \bibnamefont
  {Camposeo}}, \bibinfo {author} {\bibfnamefont {Pompilio~Del}\ \bibnamefont
  {Carro}}, \bibinfo {author} {\bibfnamefont {Vito}\ \bibnamefont {Fasano}},
  \bibinfo {author} {\bibfnamefont {Maria}\ \bibnamefont {Moffa}}, \bibinfo
  {author} {\bibfnamefont {Rita}\ \bibnamefont {Manco}}, \bibinfo {author}
  {\bibfnamefont {Stefania}\ \bibnamefont {D'Agostino}}, \ and\ \bibinfo
  {author} {\bibfnamefont {Dario}\ \bibnamefont {Pisignano}},\ }\bibfield
  {title} {\enquote {\bibinfo {title} {{Lasers: Distributed Feedback Imprinted
  Electrospun Fiber Lasers}},}\ }\href {\doibase 10.1002/adma.201470263}
  {\bibfield  {journal} {\bibinfo  {journal} {Advanced Materials}\ }\textbf
  {\bibinfo {volume} {26}},\ \bibinfo {pages} {6542--6547} (\bibinfo {year}
  {2014})}\BibitemShut {NoStop}%
\bibitem [{\citenamefont {Moran-Mirabal}\ \emph {et~al.}(2007)\citenamefont
  {Moran-Mirabal}, \citenamefont {Slinker}, \citenamefont {DeFranco},
  \citenamefont {Verbridge}, \citenamefont {Ilic}, \citenamefont
  {Flores-Torres}, \citenamefont {Abru\~{n}a}, \citenamefont {Malliaras},\ and\
  \citenamefont {Craighead}}]{Moran2007}%
  \BibitemOpen
  \bibfield  {author} {\bibinfo {author} {\bibfnamefont {Jos\'{e}~M}\
  \bibnamefont {Moran-Mirabal}}, \bibinfo {author} {\bibfnamefont {Jason~D}\
  \bibnamefont {Slinker}}, \bibinfo {author} {\bibfnamefont {John~a}\
  \bibnamefont {DeFranco}}, \bibinfo {author} {\bibfnamefont {Scott~S}\
  \bibnamefont {Verbridge}}, \bibinfo {author} {\bibfnamefont {Rob}\
  \bibnamefont {Ilic}}, \bibinfo {author} {\bibfnamefont {Samuel}\ \bibnamefont
  {Flores-Torres}}, \bibinfo {author} {\bibfnamefont {H\'{e}ctor}\ \bibnamefont
  {Abru\~{n}a}}, \bibinfo {author} {\bibfnamefont {George~G}\ \bibnamefont
  {Malliaras}}, \ and\ \bibinfo {author} {\bibfnamefont {H~G}\ \bibnamefont
  {Craighead}},\ }\bibfield  {title} {\enquote {\bibinfo {title} {{Electrospun
  light-emitting nanofibers.}}}\ }\href {\doibase 10.1021/nl062778+} {\bibfield
   {journal} {\bibinfo  {journal} {Nano letters}\ }\textbf {\bibinfo {volume}
  {7}},\ \bibinfo {pages} {458--63} (\bibinfo {year} {2007})}\BibitemShut
  {NoStop}%
\bibitem [{\citenamefont {Vohra}\ \emph {et~al.}(2011)\citenamefont {Vohra},
  \citenamefont {Giovanella}, \citenamefont {Tubino}, \citenamefont {Murata},\
  and\ \citenamefont {Botta}}]{Vohra2011}%
  \BibitemOpen
  \bibfield  {author} {\bibinfo {author} {\bibfnamefont {Varun}\ \bibnamefont
  {Vohra}}, \bibinfo {author} {\bibfnamefont {Umberto}\ \bibnamefont
  {Giovanella}}, \bibinfo {author} {\bibfnamefont {Riccardo}\ \bibnamefont
  {Tubino}}, \bibinfo {author} {\bibfnamefont {Hideyuki}\ \bibnamefont
  {Murata}}, \ and\ \bibinfo {author} {\bibfnamefont {Chiara}\ \bibnamefont
  {Botta}},\ }\bibfield  {title} {\enquote {\bibinfo {title}
  {{Electroluminescence from conjugated polymer electrospun nanofibers in
  solution processable organic light-emitting diodes.}}}\ }\href {\doibase
  10.1021/nn201029c} {\bibfield  {journal} {\bibinfo  {journal} {ACS nano}\
  }\textbf {\bibinfo {volume} {5}},\ \bibinfo {pages} {5572--8} (\bibinfo
  {year} {2011})}\BibitemShut {NoStop}%
\bibitem [{\citenamefont {Kakade}\ \emph {et~al.}(2007)\citenamefont {Kakade},
  \citenamefont {Givens}, \citenamefont {Gardner}, \citenamefont {Lee},
  \citenamefont {Chase},\ and\ \citenamefont {Rabolt}}]{Kakade2007}%
  \BibitemOpen
  \bibfield  {author} {\bibinfo {author} {\bibfnamefont {Meghana~V}\
  \bibnamefont {Kakade}}, \bibinfo {author} {\bibfnamefont {Steven}\
  \bibnamefont {Givens}}, \bibinfo {author} {\bibfnamefont {Kenncorwin}\
  \bibnamefont {Gardner}}, \bibinfo {author} {\bibfnamefont {Keun~Hyung}\
  \bibnamefont {Lee}}, \bibinfo {author} {\bibfnamefont {D~Bruce}\ \bibnamefont
  {Chase}}, \ and\ \bibinfo {author} {\bibfnamefont {John~F}\ \bibnamefont
  {Rabolt}},\ }\bibfield  {title} {\enquote {\bibinfo {title} {{Electric field
  induced orientation of polymer chains in macroscopically aligned electrospun
  polymer nanofibers.}}}\ }\href {\doibase 10.1021/ja065043f} {\bibfield
  {journal} {\bibinfo  {journal} {Journal of the American Chemical Society}\
  }\textbf {\bibinfo {volume} {129}},\ \bibinfo {pages} {2777--82} (\bibinfo
  {year} {2007})}\BibitemShut {NoStop}%
\bibitem [{\citenamefont {Pagliara}\ \emph {et~al.}(2011)\citenamefont
  {Pagliara}, \citenamefont {Vitiello}, \citenamefont {Camposeo}, \citenamefont
  {Polini}, \citenamefont {Cingolani}, \citenamefont {Scamarcio},\ and\
  \citenamefont {Pisignano}}]{Pagliara2011}%
  \BibitemOpen
  \bibfield  {author} {\bibinfo {author} {\bibfnamefont {Stefano}\ \bibnamefont
  {Pagliara}}, \bibinfo {author} {\bibfnamefont {Miriam~S.}\ \bibnamefont
  {Vitiello}}, \bibinfo {author} {\bibfnamefont {Andrea}\ \bibnamefont
  {Camposeo}}, \bibinfo {author} {\bibfnamefont {Alessandro}\ \bibnamefont
  {Polini}}, \bibinfo {author} {\bibfnamefont {Roberto}\ \bibnamefont
  {Cingolani}}, \bibinfo {author} {\bibfnamefont {Gaetano}\ \bibnamefont
  {Scamarcio}}, \ and\ \bibinfo {author} {\bibfnamefont {Dario}\ \bibnamefont
  {Pisignano}},\ }\bibfield  {title} {\enquote {\bibinfo {title} {{Optical
  Anisotropy in Single Light-Emitting Polymer Nanofibers}},}\ }\href {\doibase
  10.1021/jp204582j} {\bibfield  {journal} {\bibinfo  {journal} {The Journal of
  Physical Chemistry C}\ }\textbf {\bibinfo {volume} {115}},\ \bibinfo {pages}
  {20399--20405} (\bibinfo {year} {2011})}\BibitemShut {NoStop}%
\bibitem [{\citenamefont {Richard-Lacroix}\ and\ \citenamefont
  {Pellerin}(2012)}]{Richard-L2012}%
  \BibitemOpen
  \bibfield  {author} {\bibinfo {author} {\bibfnamefont {Marie}\ \bibnamefont
  {Richard-Lacroix}}\ and\ \bibinfo {author} {\bibfnamefont {Christian}\
  \bibnamefont {Pellerin}},\ }\bibfield  {title} {\enquote {\bibinfo {title}
  {{Orientation and Structure of Single Electrospun Nanofibers of Poly(ethylene
  terephthalate) by Confocal Raman Spectroscopy}},}\ }\href {\doibase
  10.1021/ma202749d} {\bibfield  {journal} {\bibinfo  {journal}
  {Macromolecules}\ }\textbf {\bibinfo {volume} {45}},\ \bibinfo {pages}
  {1946--1953} (\bibinfo {year} {2012})}\BibitemShut {NoStop}%
\bibitem [{\citenamefont {Pagliara}\ \emph {et~al.}(2009)\citenamefont
  {Pagliara}, \citenamefont {Camposeo}, \citenamefont {Polini}, \citenamefont
  {Cingolani},\ and\ \citenamefont {Pisignano}}]{Pagliara2009}%
  \BibitemOpen
  \bibfield  {author} {\bibinfo {author} {\bibfnamefont {Stefano}\ \bibnamefont
  {Pagliara}}, \bibinfo {author} {\bibfnamefont {Andrea}\ \bibnamefont
  {Camposeo}}, \bibinfo {author} {\bibfnamefont {Alessandro}\ \bibnamefont
  {Polini}}, \bibinfo {author} {\bibfnamefont {Roberto}\ \bibnamefont
  {Cingolani}}, \ and\ \bibinfo {author} {\bibfnamefont {Dario}\ \bibnamefont
  {Pisignano}},\ }\bibfield  {title} {\enquote {\bibinfo {title} {{Electrospun
  light-emitting nanofibers as excitation source in microfluidic devices.}}}\
  }\href {\doibase 10.1039/b906188f} {\bibfield  {journal} {\bibinfo  {journal}
  {Lab on a chip}\ }\textbf {\bibinfo {volume} {9}},\ \bibinfo {pages}
  {2851--6} (\bibinfo {year} {2009})}\BibitemShut {NoStop}%
\bibitem [{\citenamefont {Xie}\ \emph {et~al.}(2009)\citenamefont {Xie},
  \citenamefont {MacEwan}, \citenamefont {Li}, \citenamefont
  {Sakiyama-Elbert},\ and\ \citenamefont {Xia}}]{Xie2009}%
  \BibitemOpen
  \bibfield  {author} {\bibinfo {author} {\bibfnamefont {Jingwei}\ \bibnamefont
  {Xie}}, \bibinfo {author} {\bibfnamefont {Matthew~R}\ \bibnamefont
  {MacEwan}}, \bibinfo {author} {\bibfnamefont {Xiaoran}\ \bibnamefont {Li}},
  \bibinfo {author} {\bibfnamefont {Shelly~E}\ \bibnamefont {Sakiyama-Elbert}},
  \ and\ \bibinfo {author} {\bibfnamefont {Younan}\ \bibnamefont {Xia}},\
  }\bibfield  {title} {\enquote {\bibinfo {title} {{Neurite outgrowth on
  nanofiber scaffolds with different orders, structures, and surface
  properties.}}}\ }\href {\doibase 10.1021/nn900070z} {\bibfield  {journal}
  {\bibinfo  {journal} {ACS nano}\ }\textbf {\bibinfo {volume} {3}},\ \bibinfo
  {pages} {1151--9} (\bibinfo {year} {2009})}\BibitemShut {NoStop}%
\bibitem [{\citenamefont {Arinstein}\ \emph {et~al.}(2007)\citenamefont
  {Arinstein}, \citenamefont {Burman}, \citenamefont {Gendelman},\ and\
  \citenamefont {Zussman}}]{Arinstein2007}%
  \BibitemOpen
  \bibfield  {author} {\bibinfo {author} {\bibfnamefont {Arkadii}\ \bibnamefont
  {Arinstein}}, \bibinfo {author} {\bibfnamefont {Michael}\ \bibnamefont
  {Burman}}, \bibinfo {author} {\bibfnamefont {Oleg}\ \bibnamefont
  {Gendelman}}, \ and\ \bibinfo {author} {\bibfnamefont {Eyal}\ \bibnamefont
  {Zussman}},\ }\bibfield  {title} {{\enquote {\bibinfo
  {title} {{Effect of supramolecular structure on polymer nanofibre
  elasticity.}}}\ }}\href {\doibase 10.1038/nnano.2006.172} {\bibfield
  {journal} {\bibinfo  {journal} {Nature nanotechnology}\ }\textbf {\bibinfo
  {volume} {2}},\ \bibinfo {pages} {59--62} (\bibinfo {year}
  {2007})}\BibitemShut {NoStop}%
\bibitem [{\citenamefont {Hohman}\ \emph
  {et~al.}(2001{\natexlab{a}})\citenamefont {Hohman}, \citenamefont {Shin},
  \citenamefont {Rutledge},\ and\ \citenamefont {Brenner}}]{Hohman2001}%
  \BibitemOpen
  \bibfield  {author} {\bibinfo {author} {\bibfnamefont {Moses~M.}\
  \bibnamefont {Hohman}}, \bibinfo {author} {\bibfnamefont {Michael}\
  \bibnamefont {Shin}}, \bibinfo {author} {\bibfnamefont {Gregory}\
  \bibnamefont {Rutledge}}, \ and\ \bibinfo {author} {\bibfnamefont
  {Michael~P.}\ \bibnamefont {Brenner}},\ }\bibfield  {title} {\enquote
  {\bibinfo {title} {{Electrospinning and electrically forced jets. I.
  Stability theory}},}\ }\href {\doibase 10.1063/1.1383791} {\bibfield
  {journal} {\bibinfo  {journal} {Physics of Fluids}\ }\textbf {\bibinfo
  {volume} {13}},\ \bibinfo {pages} {2201} (\bibinfo {year}
  {2001}{\natexlab{a}})}\BibitemShut {NoStop}%
\bibitem [{\citenamefont {Hohman}\ \emph
  {et~al.}(2001{\natexlab{b}})\citenamefont {Hohman}, \citenamefont {Shin},
  \citenamefont {Rutledge},\ and\ \citenamefont {Brenner}}]{Hohman2001a}%
  \BibitemOpen
  \bibfield  {author} {\bibinfo {author} {\bibfnamefont {Moses~M.}\
  \bibnamefont {Hohman}}, \bibinfo {author} {\bibfnamefont {Michael}\
  \bibnamefont {Shin}}, \bibinfo {author} {\bibfnamefont {Gregory}\
  \bibnamefont {Rutledge}}, \ and\ \bibinfo {author} {\bibfnamefont
  {Michael~P.}\ \bibnamefont {Brenner}},\ }\bibfield  {title} {\enquote
  {\bibinfo {title} {{Electrospinning and electrically forced jets. II.
  Applications}},}\ }\href {\doibase 10.1063/1.1384013} {\bibfield  {journal}
  {\bibinfo  {journal} {Physics of Fluids}\ }\textbf {\bibinfo {volume} {13}},\
  \bibinfo {pages} {2221} (\bibinfo {year} {2001}{\natexlab{b}})}\BibitemShut
  {NoStop}%
\bibitem [{\citenamefont {Greenfeld}\ \emph {et~al.}(2011)\citenamefont
  {Greenfeld}, \citenamefont {Arinstein}, \citenamefont {Fezzaa}, \citenamefont
  {Rafailovich},\ and\ \citenamefont {Zussman}}]{Greenfeld2011}%
  \BibitemOpen
  \bibfield  {author} {\bibinfo {author} {\bibfnamefont {Israel}\ \bibnamefont
  {Greenfeld}}, \bibinfo {author} {\bibfnamefont {Arkadii}\ \bibnamefont
  {Arinstein}}, \bibinfo {author} {\bibfnamefont {Kamel}\ \bibnamefont
  {Fezzaa}}, \bibinfo {author} {\bibfnamefont {Miriam~H.}\ \bibnamefont
  {Rafailovich}}, \ and\ \bibinfo {author} {\bibfnamefont {Eyal}\ \bibnamefont
  {Zussman}},\ }\bibfield  {title} {\enquote {\bibinfo {title} {{Polymer
  dynamics in semidilute solution during electrospinning: A simple model and
  experimental observations}},}\ }\href {\doibase 10.1103/PhysRevE.84.041806}
  {\bibfield  {journal} {\bibinfo  {journal} {Physical Review E}\ }\textbf
  {\bibinfo {volume} {84}},\ \bibinfo {pages} {041806} (\bibinfo {year}
  {2011})}\BibitemShut {NoStop}%
\bibitem [{\citenamefont {Pontrelli}\ \emph {et~al.}(2014)\citenamefont
  {Pontrelli}, \citenamefont {Gentili}, \citenamefont {Coluzza}, \citenamefont
  {Pisignano},\ and\ \citenamefont {Succi}}]{Pontrelli}%
  \BibitemOpen
  \bibfield  {author} {\bibinfo {author} {\bibfnamefont {Giuseppe}\
  \bibnamefont {Pontrelli}}, \bibinfo {author} {\bibfnamefont {Daniele}\
  \bibnamefont {Gentili}}, \bibinfo {author} {\bibfnamefont {Ivan}\
  \bibnamefont {Coluzza}}, \bibinfo {author} {\bibfnamefont {Dario}\
  \bibnamefont {Pisignano}}, \ and\ \bibinfo {author} {\bibfnamefont {Sauro}\
  \bibnamefont {Succi}},\ }\bibfield  {title} {\enquote {\bibinfo {title}
  {{Effects of non-linear rheology on electrospinning process: A model
  study}},}\ }\href {\doibase 10.1016/j.mechrescom.2014.07.003} {\bibfield
  {journal} {\bibinfo  {journal} {Mechanics Research Communications}\ }\textbf
  {\bibinfo {volume} {61}},\ \bibinfo {pages} {41--46} (\bibinfo {year}
  {2014})},\ \Eprint {http://arxiv.org/abs/1405.6075} {arXiv:1405.6075}
  \BibitemShut {NoStop}%
\bibitem [{\citenamefont {Sun}\ \emph {et~al.}(2006)\citenamefont {Sun},
  \citenamefont {Chang}, \citenamefont {Li},\ and\ \citenamefont
  {Lin}}]{Sun2006}%
  \BibitemOpen
  \bibfield  {author} {\bibinfo {author} {\bibfnamefont {Daoheng}\ \bibnamefont
  {Sun}}, \bibinfo {author} {\bibfnamefont {Chieh}\ \bibnamefont {Chang}},
  \bibinfo {author} {\bibfnamefont {Sha}\ \bibnamefont {Li}}, \ and\ \bibinfo
  {author} {\bibfnamefont {Liwei}\ \bibnamefont {Lin}},\ }\bibfield  {title}
  {\enquote {\bibinfo {title} {{Near-field electrospinning.}}}\ }\href
  {\doibase 10.1021/nl0602701} {\bibfield  {journal} {\bibinfo  {journal} {Nano
  letters}\ }\textbf {\bibinfo {volume} {6}},\ \bibinfo {pages} {839--42}
  (\bibinfo {year} {2006})}\BibitemShut {NoStop}%
\bibitem [{\citenamefont {Bisht}\ \emph {et~al.}(2011)\citenamefont {Bisht},
  \citenamefont {Canton}, \citenamefont {Mirsepassi}, \citenamefont {Kulinsky},
  \citenamefont {Oh}, \citenamefont {Dunn-Rankin},\ and\ \citenamefont
  {Madou}}]{Bisht2011}%
  \BibitemOpen
  \bibfield  {author} {\bibinfo {author} {\bibfnamefont {Gobind~S}\
  \bibnamefont {Bisht}}, \bibinfo {author} {\bibfnamefont {Giulia}\
  \bibnamefont {Canton}}, \bibinfo {author} {\bibfnamefont {Alireza}\
  \bibnamefont {Mirsepassi}}, \bibinfo {author} {\bibfnamefont {Lawrence}\
  \bibnamefont {Kulinsky}}, \bibinfo {author} {\bibfnamefont {Seajin}\
  \bibnamefont {Oh}}, \bibinfo {author} {\bibfnamefont {Derek}\ \bibnamefont
  {Dunn-Rankin}}, \ and\ \bibinfo {author} {\bibfnamefont {Marc~J}\
  \bibnamefont {Madou}},\ }\bibfield  {title} {\enquote {\bibinfo {title}
  {{Controlled continuous patterning of polymeric nanofibers on
  three-dimensional substrates using low-voltage near-field
  electrospinning.}}}\ }\href {\doibase 10.1021/nl2006164} {\bibfield
  {journal} {\bibinfo  {journal} {Nano letters}\ }\textbf {\bibinfo {volume}
  {11}},\ \bibinfo {pages} {1831--7} (\bibinfo {year} {2011})}\BibitemShut
  {NoStop}%
\bibitem [{\citenamefont {Pai}\ \emph {et~al.}(2011)\citenamefont {Pai},
  \citenamefont {Boyce},\ and\ \citenamefont {Rutledge}}]{Pai2011}%
  \BibitemOpen
  \bibfield  {author} {\bibinfo {author} {\bibfnamefont {Chia-Ling}\
  \bibnamefont {Pai}}, \bibinfo {author} {\bibfnamefont {Mary~C.}\ \bibnamefont
  {Boyce}}, \ and\ \bibinfo {author} {\bibfnamefont {Gregory~C.}\ \bibnamefont
  {Rutledge}},\ }\bibfield  {title} {\enquote {\bibinfo {title} {{Mechanical
  properties of individual electrospun PA 6(3)T fibers and their variation with
  fiber diameter}},}\ }\href {\doibase 10.1016/j.polymer.2011.03.041}
  {\bibfield  {journal} {\bibinfo  {journal} {Polymer}\ }\textbf {\bibinfo
  {volume} {52}},\ \bibinfo {pages} {2295--2301} (\bibinfo {year}
  {2011})}\BibitemShut {NoStop}%
\bibitem [{\citenamefont {Yarin}(1990)}]{Yarin1990}%
  \BibitemOpen
  \bibfield  {author} {\bibinfo {author} {\bibfnamefont {Alexander~L.}\
  \bibnamefont {Yarin}},\ }\bibfield  {title} {\enquote {\bibinfo {title}
  {{Strong flows of polymeric liquids}},}\ }\href@noop {} {\bibfield  {journal}
  {\bibinfo  {journal} {Journal of Non-Newtonian Fluid Mechanics}\ }\textbf
  {\bibinfo {volume} {37}},\ \bibinfo {pages} {113--138} (\bibinfo {year}
  {1990})}\BibitemShut {NoStop}%
\bibitem [{\citenamefont {Reneker}\ and\ \citenamefont
  {Yarin}(2008)}]{Reneker2008}%
  \BibitemOpen
  \bibfield  {author} {\bibinfo {author} {\bibfnamefont {Darrell~H.}\
  \bibnamefont {Reneker}}\ and\ \bibinfo {author} {\bibfnamefont
  {Alexander~L.}\ \bibnamefont {Yarin}},\ }\bibfield  {title} {\enquote
  {\bibinfo {title} {{Electrospinning jets and polymer nanofibers}},}\ }\href
  {\doibase 10.1016/j.polymer.2008.02.002} {\bibfield  {journal} {\bibinfo
  {journal} {Polymer}\ }\textbf {\bibinfo {volume} {49}},\ \bibinfo {pages}
  {2387--2425} (\bibinfo {year} {2008})}\BibitemShut {NoStop}%
\bibitem [{\citenamefont {Bhattacharjee}\ and\ \citenamefont
  {Rutledge}(2011)}]{Bhattacharjee2011}%
  \BibitemOpen
  \bibfield  {author} {\bibinfo {author} {\bibfnamefont {P.~K.}\ \bibnamefont
  {Bhattacharjee}}\ and\ \bibinfo {author} {\bibfnamefont {G.~C.}\ \bibnamefont
  {Rutledge}},\ }\href {\doibase 10.1016/B978-0-08-055294-1.00039-8} {\emph
  {\bibinfo {title} {{Comprehensive Biomaterials}}}},\ edited by\ \bibinfo
  {editor} {\bibfnamefont {P.}~\bibnamefont {Ducheyne}}, \bibinfo {editor}
  {\bibfnamefont {K.~E.}\ \bibnamefont {Healy}}, \bibinfo {editor}
  {\bibfnamefont {D.~W.}\ \bibnamefont {Hutmacher}}, \bibinfo {editor}
  {\bibfnamefont {D.~W.}\ \bibnamefont {Grainger}}, \ and\ \bibinfo {editor}
  {\bibfnamefont {C.~J.}\ \bibnamefont {Kirkpatrick}}\ (\bibinfo  {publisher}
  {Elsevier},\ \bibinfo {address} {Oxford},\ \bibinfo {year}
  {2011})\BibitemShut {NoStop}%
\bibitem [{\citenamefont {Chiu-Webster}\ and\ \citenamefont
  {Lister}(2006)}]{Chiu-Webster2006}%
  \BibitemOpen
  \bibfield  {author} {\bibinfo {author} {\bibfnamefont {S.}~\bibnamefont
  {Chiu-Webster}}\ and\ \bibinfo {author} {\bibfnamefont {J.~R.}\ \bibnamefont
  {Lister}},\ }\bibfield  {title} {\enquote {\bibinfo {title} {{The fall of a
  viscous thread onto a moving surface: a ‘fluid-mechanical sewing
  machine’}},}\ }\href {\doibase 10.1017/S0022112006002503} {\bibfield
  {journal} {\bibinfo  {journal} {Journal of Fluid Mechanics}\ }\textbf
  {\bibinfo {volume} {569}},\ \bibinfo {pages} {89} (\bibinfo {year}
  {2006})}\BibitemShut {NoStop}%
\bibitem [{\citenamefont {Li}\ \emph {et~al.}(2013)\citenamefont {Li},
  \citenamefont {Ga\~{n}\'{a}n Calvo}, \citenamefont {L\'{o}pez-Herrera},
  \citenamefont {Yin},\ and\ \citenamefont {Yin}}]{Li2013}%
  \BibitemOpen
  \bibfield  {author} {\bibinfo {author} {\bibfnamefont {Fang}\ \bibnamefont
  {Li}}, \bibinfo {author} {\bibfnamefont {Alfonso~M.}\ \bibnamefont
  {Ga\~{n}\'{a}n Calvo}}, \bibinfo {author} {\bibfnamefont {Jos\'{e}~M.}\
  \bibnamefont {L\'{o}pez-Herrera}}, \bibinfo {author} {\bibfnamefont
  {Xie-Yuan}\ \bibnamefont {Yin}}, \ and\ \bibinfo {author} {\bibfnamefont
  {Xie-Zhen}\ \bibnamefont {Yin}},\ }\bibfield  {title} {\enquote {\bibinfo
  {title} {{Absolute and convective instability of a charged viscoelastic
  liquid jet}},}\ }\href {\doibase 10.1016/j.jnnfm.2013.01.003} {\bibfield
  {journal} {\bibinfo  {journal} {Journal of Non-Newtonian Fluid Mechanics}\
  }\textbf {\bibinfo {volume} {196}},\ \bibinfo {pages} {58--69} (\bibinfo
  {year} {2013})}\BibitemShut {NoStop}%
\bibitem [{\citenamefont {Li}\ \emph {et~al.}(2006)\citenamefont {Li},
  \citenamefont {Yin},\ and\ \citenamefont {Yin}}]{Li2006}%
  \BibitemOpen
  \bibfield  {author} {\bibinfo {author} {\bibfnamefont {Fang}\ \bibnamefont
  {Li}}, \bibinfo {author} {\bibfnamefont {Xie-Yuan}\ \bibnamefont {Yin}}, \
  and\ \bibinfo {author} {\bibfnamefont {Xie-Zhen}\ \bibnamefont {Yin}},\
  }\bibfield  {title} {\enquote {\bibinfo {title} {{Instability analysis of a
  coaxial jet under a radial electric field in the nonequipotential case}},}\
  }\href {\doibase 10.1063/1.2181604} {\bibfield  {journal} {\bibinfo
  {journal} {Physics of Fluids}\ }\textbf {\bibinfo {volume} {18}},\ \bibinfo
  {pages} {037101} (\bibinfo {year} {2006})}\BibitemShut {NoStop}%
\end{thebibliography}%

\end{document}